\documentclass{ieeeaccess}
\usepackage{cite}
\usepackage{amsmath,amssymb,amsfonts}
\usepackage{algorithmic}
\usepackage{graphicx}
\usepackage{textcomp}
\usepackage{hyperref}
\usepackage{multirow}
\usepackage{array}
\usepackage{makecell}
\usepackage{tabularx,booktabs}
\usepackage{xtab,afterpage}

\def\BibTeX{{\rm B\kern-.05em{\sc i\kern-.025em b}\kern-.08em
    T\kern-.1667em\lower.7ex\hbox{E}\kern-.125emX}}
\begin{document}
\history{Date of publication xxxx 00, 0000, date of current version xxxx 00, 0000.}
\doi{10.1109/ACCESS.2017.DOI}

\title{Multilingual Audio-Visual Smartphone Dataset And Evaluation}
\author{\uppercase{Hareesh Mandalapu}\authorrefmark{1},
\uppercase{Aravinda Reddy P N}\authorrefmark{2}, \uppercase{Raghavendra Ramachandra}\authorrefmark{1}, \IEEEmembership{Senior Member, IEEE}, \uppercase{K Sreenivasa  Rao}\authorrefmark{3}, \IEEEmembership{Member, IEEE}, \uppercase{Pabitra Mitra} \authorrefmark{3}, \IEEEmembership{Member, IEEE}, and \uppercase{S R Mahadeva Prasanna} \authorrefmark{4}, \IEEEmembership{Member, IEEE}, \uppercase{Christoph Busch}\authorrefmark{1}, \IEEEmembership{Senior Member, IEEE}}
\address[1]{Department of Information Security and Communication Technology, Norwegian University of Science and Technology (NTNU), Gjøvik, 2815, Norway}
\address[2]{Advanced Technology Development Centre, Indian Institute of Technology, Kharagpur, West Bengal, 721302, India.}
\address[3]{Department of Computer Science and Engineering, Indian Institute of Technology, Kharagpur, West Bengal, 721302, India.}
\address[4]{Department  of  Electrical  Engineering, Indian Institute of Technology, Dharwad, Karnataka, 580011.  India.}


\tfootnote{This work was supported by Department of Information Security and Communication Technology, NTNU, Gjøvik and research Council of Norway under Grant IKTPLUSS 248030/O70 and advanced Technology Development Centre, Indian Institute of Technology, Kharagpur, India.}

\markboth
{H.Mandalapu  \headeretal: Multilingual Audio-Visual Smartphone Dataset And Evaluation}
{H.Mandalapu  \headeretal: Multilingual Audio-Visual Smartphone Dataset And Evaluation}

\corresp{Corresponding author: Hareesh Mandalapu (e-mail: hareesh.mandalapu@ntnu.no).}

\begin{abstract}
Smartphones have been employed with biometric-based verification systems to provide security in highly sensitive applications. Audio-visual biometrics are getting popular due to their usability, and also it will be challenging to spoof because of their multimodal nature. In this work, we present an audio-visual smartphone dataset captured in five different recent smartphones. This new dataset contains 103 subjects captured in three different sessions considering the different real-world scenarios. Three different languages are acquired in this dataset to include the problem of language dependency of the speaker recognition systems. These unique characteristics of this dataset will pave the way to implement novel state-of-the-art unimodal or audio-visual speaker recognition systems. We also report the performance of the bench-marked biometric verification systems on our dataset. The robustness of biometric algorithms is evaluated towards multiple dependencies like signal noise, device, language and presentation attacks like replay and synthesized signals with extensive experiments. The obtained results raised many concerns about the generalization properties of state-of-the-art biometrics methods in smartphones. 

\end{abstract}


\titlepgskip=-15pt

\maketitle

\section{Introduction} \label{sec:introduction}

With the advances in biometrics, the usage of passwords and smart cards to gain access into several control applications have been slowly depreciated. Henceforth for reliable and secure access control, biometrics have been deployed in various applications, including smartphone unlocking, banking transactions, financial services, border control, etc. The biometrics in access control applications improve trustworthiness and enhance user proficiency by verifying who they are. A biometric system aims to recognize the person based on their physiological or behavioural characteristics based on ISO/IEC 2382-37. The physiological characteristics include the face, iris, fingerprint etc., and behavioural characteristics include speech, keystroke, gait etc. 

Smartphone biometrics has grown expeditiously over the years. The number of smartphone users crossed 3 billion in 2020 and is expected to increase in millions in the coming years. According to the Mercator Advisory Group report, 66\% of smartphone users are expected to use biometrics for authentication by the end of 2024. In 2020, 41\% of smartphone users used biometrics which was 27\% in 2019. Among different biometric modalities, fingerprint-based authentication is at the top. However, the amount of users for face and biometrics has been increasing. Voice-based recognition increased to 20\% in 2020, from 11\% in 2019 and face recognition jumped to 30\% in 2020, from 20\% in 2019. The application of smartphone biometrics has been widely used in mobile banking, e-commerce, remote identification etc.

Different types of smartphones like Android, iPhone and blackberry provide uni-modal applications based on either fingerprint, iris or face recognition, and recently speech has been added as a biometric cue for authentication purposes. The built-in biometrics are not fixed for all smartphones. For example, some smartphones come with fingerprint, and some include face recognition. The captured uni-modal biometrics like face or iris comes with several problems like low quality, variations in pose, problem with illuminations, background noise, low spatial and temporal resolutions of video \cite{mandalapu2021audio}. Therefore, this problem is addressed in multimodal biometrics by taking advantage of default sensors like cameras and microphones. Multimodal systems like audio-visual biometrics utilize the complementary information of face and speech and exploit the user-friendly capture of face and voice in a single recording. Audio-visual biometric data capture is cost-effective and can be carried out without additional sensors (e.g., fingerprint reader or iris camera).

The applications based on biometrics in smartphones has several advantages but also exist several challenges. The key challenges are the robustness and generalizability of a biometric system caused by algorithm dependencies and evolving presentation attacks. The aforementioned challenges are the main problems that circumscribe reliable and secure smartphone-based applications. The first challenge is the algorithm dependencies which limits the interoperability of a biometric algorithm across multiple types of smartphones. Interoperability is defined as the ability of a biometric system to handle variations introduced in the biometric data due to different capture devices. Due to different kinds of smartphone sensors, capturing conditions and human behaviour. The dependency of the biometric algorithm on particular data properties limits the robustness of optimal recognition. Therefore, it is very challenging to develop a conventional biometric method for a wide variety of smartphones. 

The second challenge is from the presentation attacks or also called spoofing attacks and indirect attacks, which are comprehensively explained in \cite{ramachandra2017presentation} for face and in \cite{mandalapu2021audio} for audio-visual. Presentation attacks are defined as the presentation to a biometric capture subsystem with the goal of interfering with the operation of the biometric system \cite{ISO-IEC-30107-1-PAD-Framework-160115}. Presentation attacks have become easy to create and use as a concealer or impostor towards the target subject. Growing presentation attacks and limitations in smartphone sensors cause major problems questioning the performance of smartphone biometrics.


The factors above motivated research on the study of smartphone biometrics towards the key challenges. In this direction, to examine the challenges, we need a smartphone biometrics database with different attributes. There are few biometric databases have been created using smartphones in both uni-modal \cite{rattani2018survey} and multimodal biometrics \cite{mccool2009mobio,ramachandra2019smartphone}. However, the existing databases are limited with several devices, languages and sessions. Therefore, we have created a multilingual audio-visual smartphone (MAVS) dataset considering smartphone devices, sessions, speech languages and presentation attacks. The novel dataset contains audio-visual biometric data of 103 subjects (70 male, 33 female) captured in three sessions with variable noise and illumination. Each subject utters six sentences, each in three different languages and recorded in five different smartphones. We have also created two types of presentation attacks in both audio, video and audio-visual scenarios. The first type of attack is a physical access attack which is created by replaying an audio-visual sample on a display-speaker setup and recorded using a smartphone. The second attack is a synthesized attack where audio and video are created separately via speech synthesis and face-swapping.

Further, we have benchmarked the dataset by performing extensive experiments in two directions. The first direction is to observe the biometric algorithm dependencies concerning device, illumination, background noise and language. The second direction is to examine the vulnerability towards presentation attacks. The baseline presentation attack detection methods in both audio and visual domains are included in this work. The biometric recognition algorithms are chosen from the state-of-the-art methods from the literature. The experimental results are presented in ISO/IEC biometric standards \cite{ISO-IEC-19795-4-071128} with pictorial representations and detailed discussion. 
 
The rest of the paper is organized as follows. Section \ref{sec:related_work} presents the related work in audio-visual datasets with sample images and discussion of results. The detailed description of the multilingual audio-visual smartphone (MAVS) dataset created in this research is presented in Section \ref{sec:dataset_details}. Section \ref{sec:perform_eval} describes the performance evaluation protocols used in bench-marking the MAVS dataset. Section \ref{sec:exp_results} presents the experiments performed and results obtained and Section \ref{sec:conclusion} concludes this paper with discussion on the future work.

\section{Related Work} \label{sec:related_work}


The sensitivity of data in smartphone utilization has made the usage of biometrics a critical feature. Therefore, the research in smartphone biometrics has obtained much attention in recent years. The built-in biometric sensors provide the necessary authentication for many smartphones. However, the inconsistency of performance in these devices encouraged a new direction of biometric recognition using the default sensors like camera and microphone. In this direction, few audio-visual smartphone biometric datasets have been developed by capturing talking subjects' videos. Multimodal biometric databases captured modalities like a finger photo, face, iris photo, and speech data. However, considering the standard sensors in all smartphones, we studied only audio-visual databases, including face and voice. In this section, we present a comprehensive study on audio-visual biometric databases. A detailed study on all audio-visual biometric databases is performed in \cite{mandalapu2021audio} by Mandalapu \textit{et al.} along with a comparison of best-performing algorithms. In this section, we present some audio-visual databases in detail.

Early audio-visual biometric datasets are created by the advanced multimedia processing (AMP) lab of Carnegie Melon University (CMU) \footnote{The AMP/CMU dataset: \url{http://amp.ece.cmu.edu/}}. With ten subjects, each speaking 78 isolated words, the recording is taken by a digital camcorder with a tie-clip microphone \cite{chen2001audiovisual}. The dataset is made publicly available with sound files and lip parameters. Although the number of subjects is low, this dataset assisted in developing a visual shape-based feature vector for audio-visual speaker recognition in \cite{aleksic2003audio}. Biometrics Access Control for Networked and E-Commerce Applications (BANCA)\footnote{The BANCA database: \url{http://www.ee.surrey.ac.uk/CVSSP/banca/}} \cite{bailly2003banca} is developed for E-Commerce applications. Important features in this database are multiple European languages captured using both high and low-quality devices under three different scenarios: controlled, degraded, and adverse. Also, the total number of subjects was 208, with an equal number of men and women. Figure \ref{fig:banca} shows the sample images of this database from three different scenarios.

\Figure[htb]()[width=0.75\columnwidth]{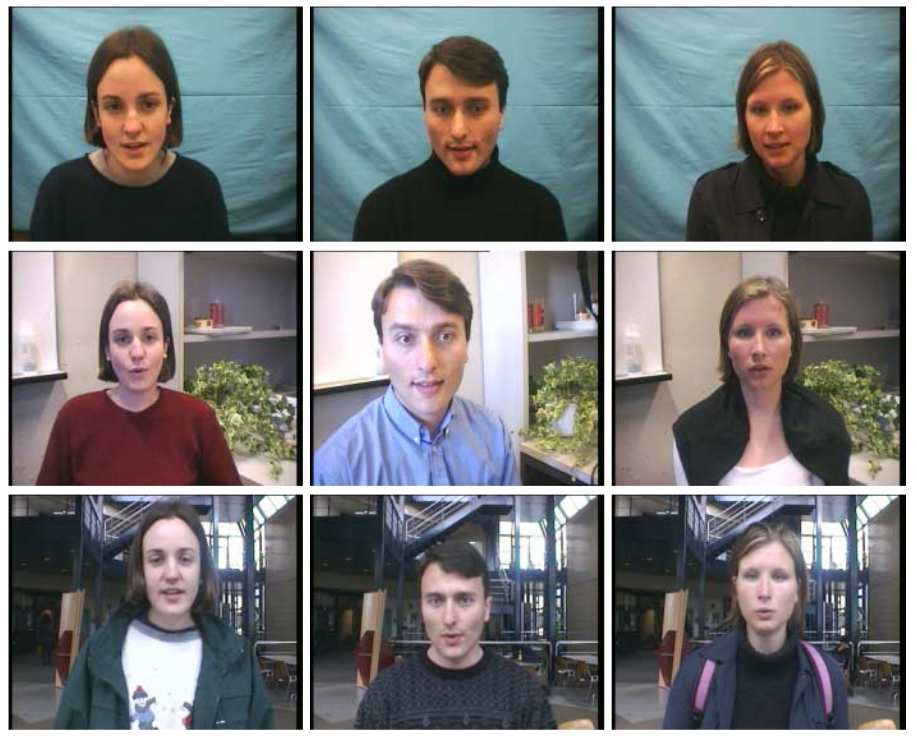}
{Example BANCA database images Up: Controlled, Middle: Degraded and Down: Adverse scenarios \cite{bailly2003banca}.\label{fig:banca}}

The goal of multimodal biometrics is to improve the robustness of the recognition/verification process. The VALID database was created in a realistic audio-visual noisy office room under uncontrolled lighting and acoustic noise. The VALID database is publicly available to research purposes \footnote{The VALID database: \url{http://ee.ucd.ie/validdb/}}. The MultiModal Verification for Teleservices and Security (M2VTS) applications database has been developed for granting access to secure regions using audio-visual person verification \cite{m2vts}. An extension to the M2VTS database is XM2VTS (extended M2VTS) with focus on high-quality biometric samples \cite{messer1999xm2vtsdb}. It contains high-quality face images, 32 kHz 16-bit audio files, video sequences, and a 3D Model. The database is publicly available at cost price \footnote{The XM2VTS database: \url{http://www. ee.surrey.ac.uk/CVSSP/xm2vtsdb/}}.

\Figure[htb]()[width=0.75\columnwidth]{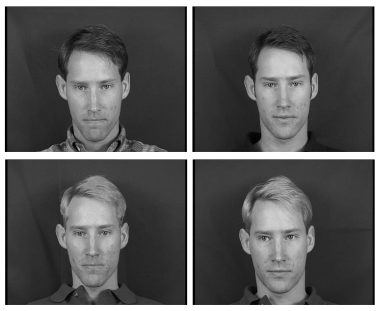}
{Front profile shots of a subject from four sessions of XM2VTS database \cite{messer1999xm2vtsdb}. \label{fig:xm2vts}}

Video recordings of people reading sentences from Texas Instruments and Massachusetts Institute of Technology (TIMIT) corpus (VidTIMID) \footnote{The VidTIMTI dataset: \url{http://conradsanderson.id.au/vidtimit/}} is a publicly available dataset presented in \cite{sanderson2009multi}. A distinctive part of VidTIMIT dataset is that it also contains head rotation sequence for each person in each session \cite{sanderson2002vidtimit}. BioSecure\footnote{BioSecure: \url{https://biosecure.wp.tem-tsp.eu/biosecure-database/}} is a popular multimodal database that also comprises of audio-visual dataset \cite{ortega2010multiscenario}. The database consists of data from 600 subjects recorded in three different scenarios. The sample images from the database are shown in Figure \ref{fig:biosecure}.

\Figure[htb]()[width=0.96\columnwidth]{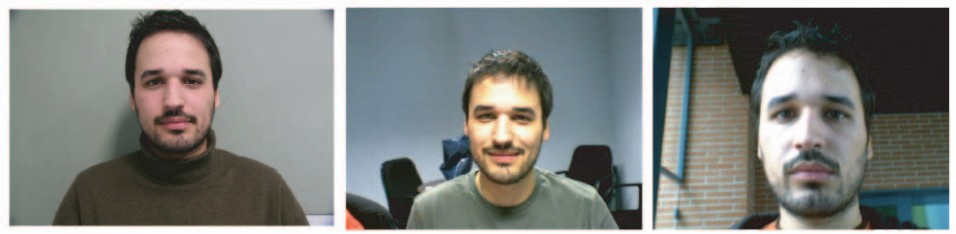}
{Face samples acquired in BioSecure database in three different scenarios. Left: indoor digital camera (from DS2), Middle: Webcam (from DS2), and Right: outdoor Webcam (from DS3) \cite{ortega2010multiscenario}. \label{fig:biosecure}}

\begin{table*}[t!]
 \centering
 \caption{Details of Audio-visual Biometric Verification Databases.}
 \label{tab:datasets}
 \begin{tabular}{|c|c|c|c|c|c|} \hline
  \multirow{2}{*}{\textbf{Dataset}} & \multirow{2}{*}{\textbf{Year}} & \multirow{2}{*}{\textbf{Devices}} & \multirow{2}{*}{\textbf{No. of subjects}} & \multirow{2}{*}{\textbf{Biometric}} & \multirow{2}{*}{\textbf{Availability}}  \\ 
  & & & & & \\ \hline

  \multirow{2}{*}{AMP/CMU \cite{chen2001audiovisual}} & \multirow{2}{*}{2001} & Digital Camcorder, & 10 & \multirow{2}{*}{Face, voice} & \multirow{2}{*}{Free} \\ 
  && tie-clip microphone&(7 M, 3 F) &  &\\ \hline
  
  \multirow{2}{*}{BANCA \cite{bailly2003banca}} & \multirow{2}{*}{2003} & Webcam and & 208 & \multirow{2}{*}{Face, voice} & \multirow{2}{*}{Free} \\ 
  && Digital Camera&(104 M, 104 F) & &\\ \hline
  
  \multirow{2}{*}{VALID \cite{fox2005valid}} & \multirow{2}{*}{2005} & Canon 3CCD XM1 & 106 & \multirow{2}{*}{Face, voice} & \multirow{2}{*}{Free} \\ 
  &&PAL &(77 M, 29 F)& & \\ \hline
  
  \multirow{2}{*}{M2VTS \cite{m2vts}} & \multirow{2}{*}{2005} & Hi8 camera, & \multirow{2}{*}{37} & \multirow{2}{*}{Face, voice} & \multirow{2}{*}{Free} \\
  && D1 digital recorder& & & \\\hline
  
  \multirow{2}{*}{XM2VTS \cite{messer1999xm2vtsdb}} & \multirow{2}{*}{2005} & Sony VX1000E, & \multirow{2}{*}{295} & \multirow{2}{*}{Face, voice} & \multirow{2}{*}{Free}\\
  && DHR1000UX & & & \\\hline
  
  \multirow{2}{*}{VidTIMIT \cite{sanderson2009multi}} & \multirow{2}{*}{2009} & Digital video &43 & \multirow{2}{*}{Face, voice} & \multirow{2}{*}{Free} \\ 
  && camera&(24 M, 19 F)& & \\ \hline
  
  \multirow{4}{*}{BioSecure \cite{ortega2010multiscenario}} & \multirow{4}{*}{2010} & Samsung Q1, & DS1: 971 & \multirow{2}{*}{Face, Fingerprint}& \multirow{4}{*}{Paid} \\ 
  & & Philips SP900NC & DS2: 667 & & \\ 
  & & HP iPAQ hx2790 & DS3: 713 & Voice, Signature& \\ 
  & & Webcam, PDA & & & \\ \hline
  
  \multirow{2}{*}{MOBIO \cite{mccool2009mobio}} & \multirow{2}{*}{2012} & Nokia N93i & \multirow{2}{*}{152} & Voice, Face & \multirow{2}{*}{Free}\\ 
  &&Mac-book&& periocular & \\ \hline
  
  \multirow{2}{*}{MobBIO \cite{sequeira2014mobbio}} & \multirow{2}{*}{2014} & Asus Transformer & \multirow{2}{*}{105} & \multirow{2}{*}{-} & \multirow{2}{*}{-} \\ 
  &&Pad TF 300T&&& \\ \hline
  
  \multirow{2}{*}{Hu \textit{et al.}\cite{hu2015deep} } & \multirow{2}{*}{2015} & \multirow{2}{*}{-} & \multirow{2}{*}{11} & \multirow{2}{*}{Audio-Visual} & \multirow{2}{*}{Free}\\ 
  & & & & &   \\ \hline


  \multirow{2}{*}{SWAN database \cite{ramachandra2019smartphone}} & \multirow{2}{*}{2019} & iPhone 6 & \multirow{2}{*}{88} & Face, Periocular, Multilingual Voice & \multirow{2}{*}{Free}  \\
   & &iPad Pro& & Presentation Attack dataset &   \\ \hline \hline
   
  \multirow{2}{*}{MAVS dataset} & \multirow{2}{*}{2021} & iPhone 6, iPhone 10, iPhone 11 & 103 & Face, Multilingual Voice & \multirow{2}{*}{Free}  \\
   & &Samsung S7 and Samsung S8 & (70 M, 33 F) & Presentation Attack dataset &   \\ \hline

 \end{tabular}
\end{table*}

The aforementioned audio-visual datasets are captured with different types of sensors. In some cases, the audio and video capturing sensors are two different devices, and the data is presented separately. However, in smartphones, the built-in camera and microphone can be used to create audio-visual data. The MOBIO database \footnote{The MOBIO database: \url{https://www.idiap.ch/dataset/mobio}} \cite{mccool2009mobio} is a audio-visual data created using a mobile phone (NOKIA N93i) and a laptop computer (2008 MacBook). MOBIO dataset helped in the study of person identification in a mobile phone environment \cite{motlicek2012bi}. In a similar fashion, the MobBIO database is developed by Sequeira \textit{et al.} in \cite{sequeira2014mobbio}. The sensors used in this work are the rear camera of the Asus Transformer Pad TF 300T. 

\Figure[!ht]()[width=0.96\columnwidth]{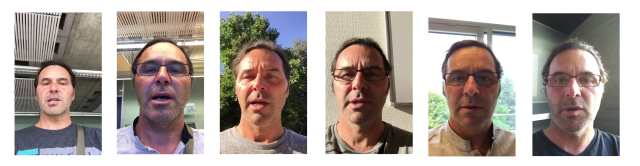}
{Talking face samples from SWAN database one frame from each session \cite{ramachandra2019smartphone}. \label{fig:swan}}

The Smartphone Multimodal Biometric database was collected for the application of mobile banking \cite{ramachandra2019smartphone}. The real-world scenarios are attributed in this database with multiple sessions and languages using iPhone 6s and iPad Pro. Along with audio-visual data, the SWAN database also contains face, eye region, finger photo and voice data. Presentation attacks are also provided as a part of this database. Figure \ref{fig:swan} shows the sample images of subjects from six sessions.

The existing databases on audio-visual biometrics provide limited variance in addressing the problem of robustness—most databases on session variance but not on device variance and language dependency. Alongside, presentation attacks are growing widely and displaying a huge impact on the optimal performance of biometric algorithms. We have formulated advanced protocols to create a multilingual audio-visual smartphone (MAVS) database considering all these problems. In this direction, the significant contributions of this paper are mentioned as follows.

\begin{enumerate}
    \item A novel multilingual audio-visual smartphone dataset will be made available for research purposes. The uniqueness of this dataset is described below.
    \begin{itemize}
        \item Biometric data from 70 male and 33 female subjects from various backgrounds.
        \item Three language speeches and three sessions (variable illumination and background noise) for all the subjects.
        \item Data recorded on multiple smartphone devices: iPhone 6s, iPhone 10, iPhone 11, Samsung S7 and Samsung S8.
        \item Three unique and three common sentences for each subject, each device, each language and each session.
        \item Two types of presentation attacks are created, each in physical access and logical access scenarios.
    \end{itemize}
    \item Benchmarking the dataset with state-of-the-art face recognition, speaker recognition algorithms and score-level fusion biometric methods.
    \item Evaluating the vulnerability of presentation attacks on state-of-the-art biometric verification and testing baseline presentation attack detection methods.
\end{enumerate}

\section{Multilingual Audio-Visual Smartphone (MAVS) Dataset} \label{sec:dataset_details}

\subsection{Acquisition}

\Figure[hbt]()[width=0.45\columnwidth]{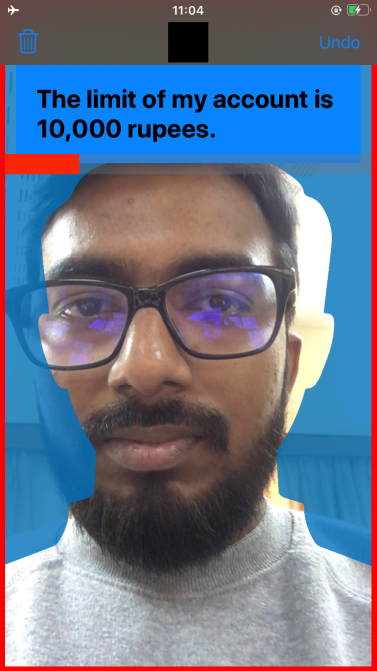}
{Mobile application (iOS) interface for data capturing.\label{fig:app_mavsd}}

\begin{figure*}[tbh]
    \centering
    \includegraphics[width=.3\columnwidth]{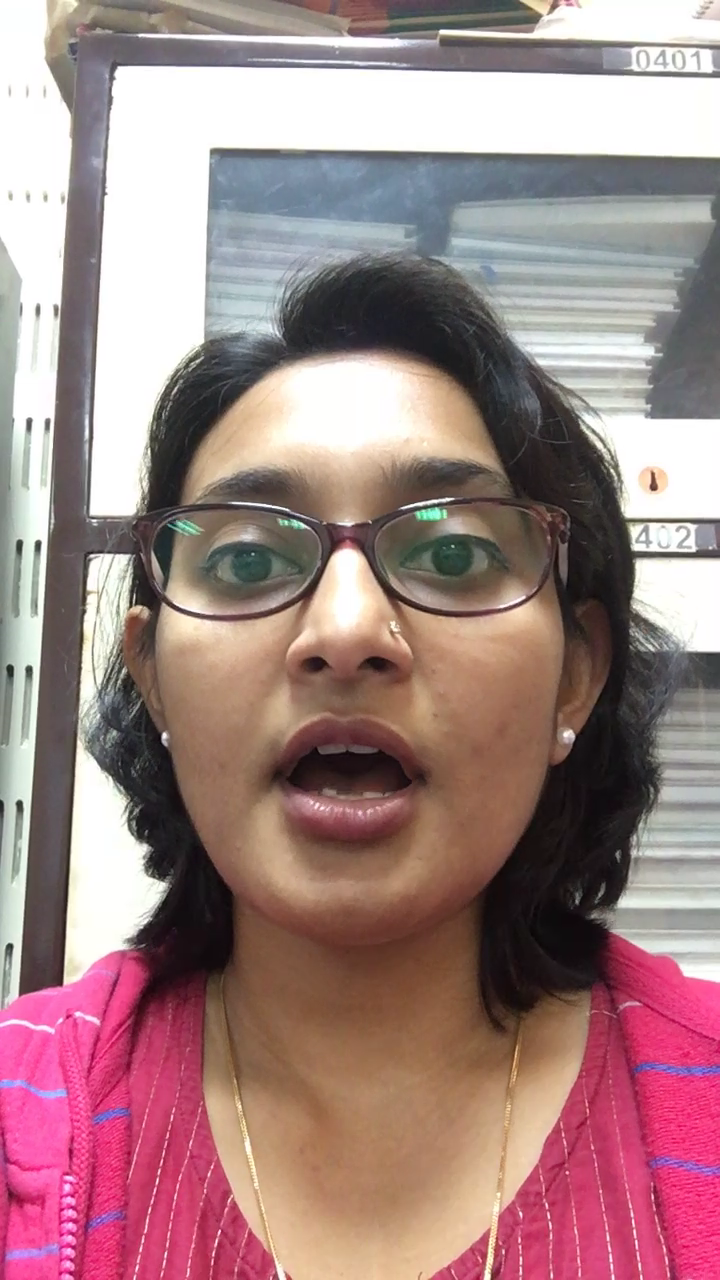}
    \includegraphics[width=.3\columnwidth]{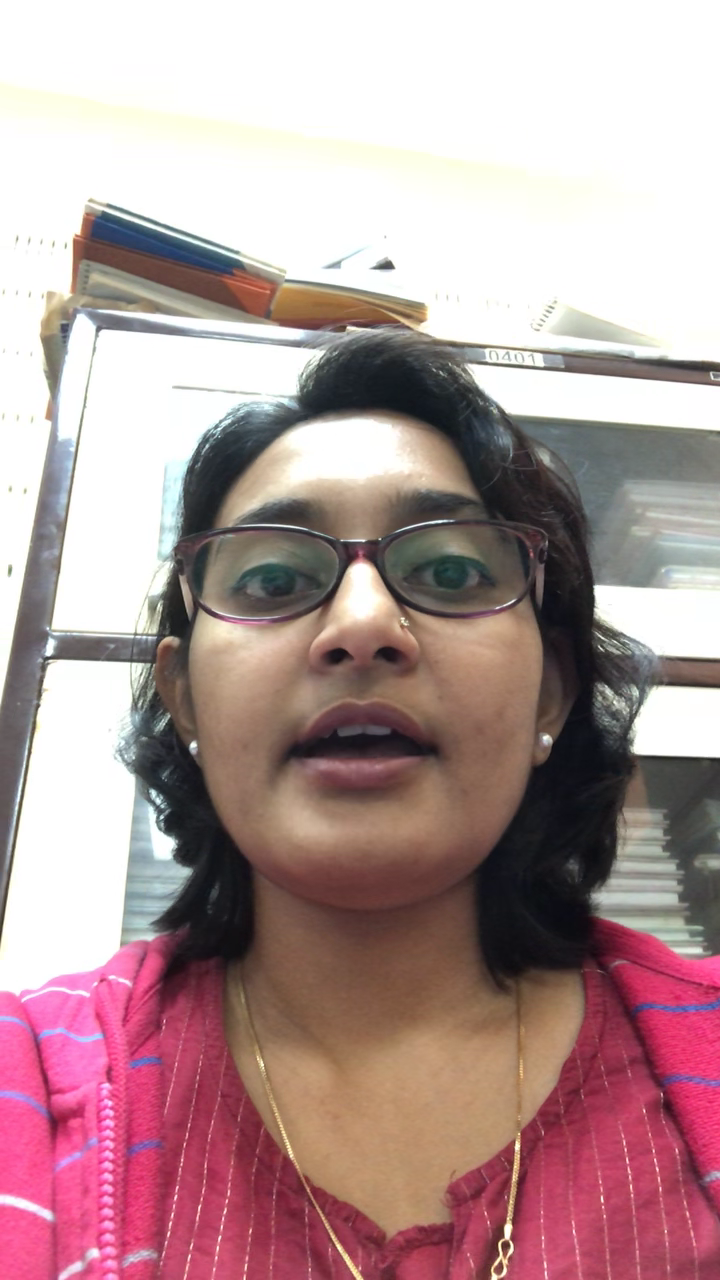}
    \includegraphics[width=.3\columnwidth]{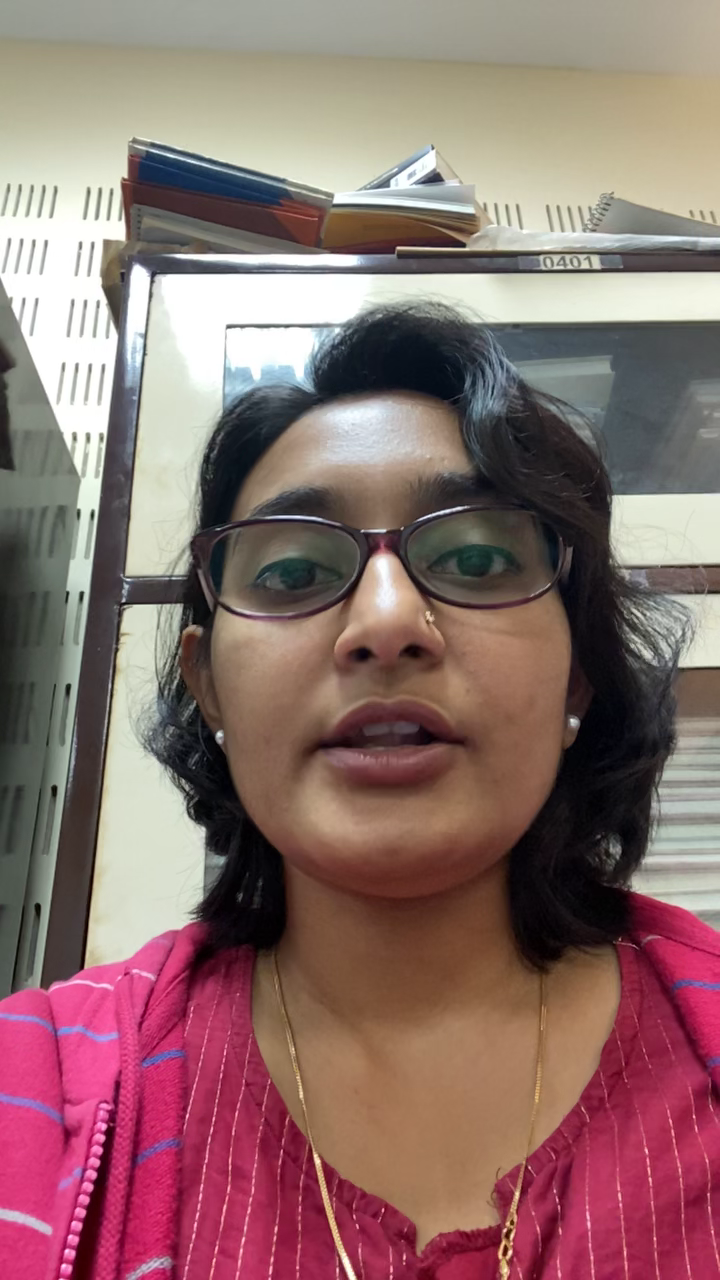}
    \includegraphics[width=.3\columnwidth]{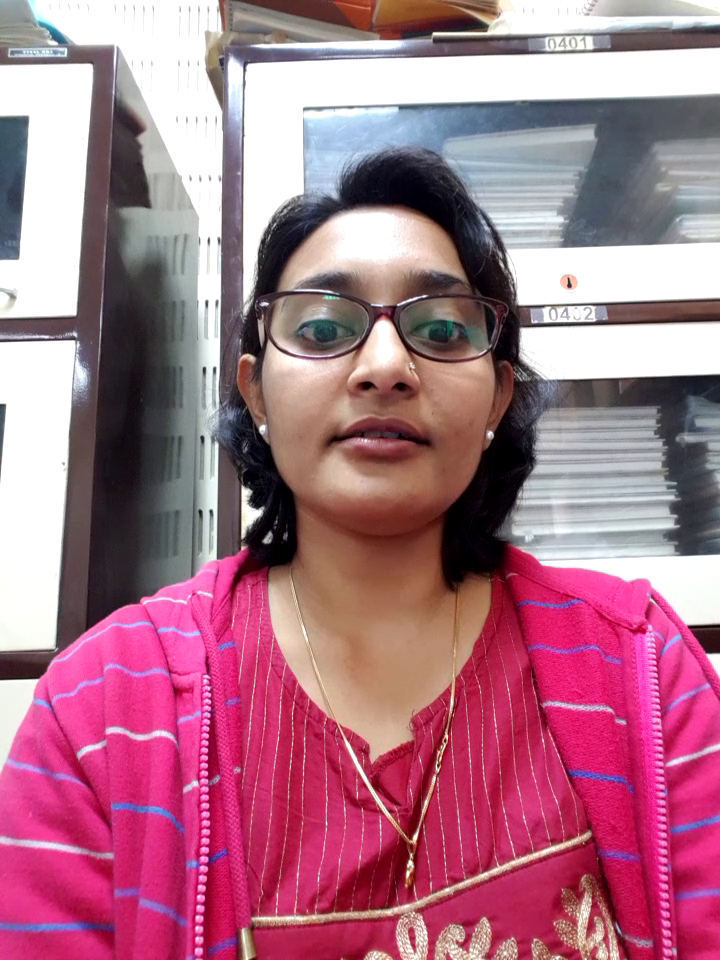}
    \includegraphics[width=.3\columnwidth]{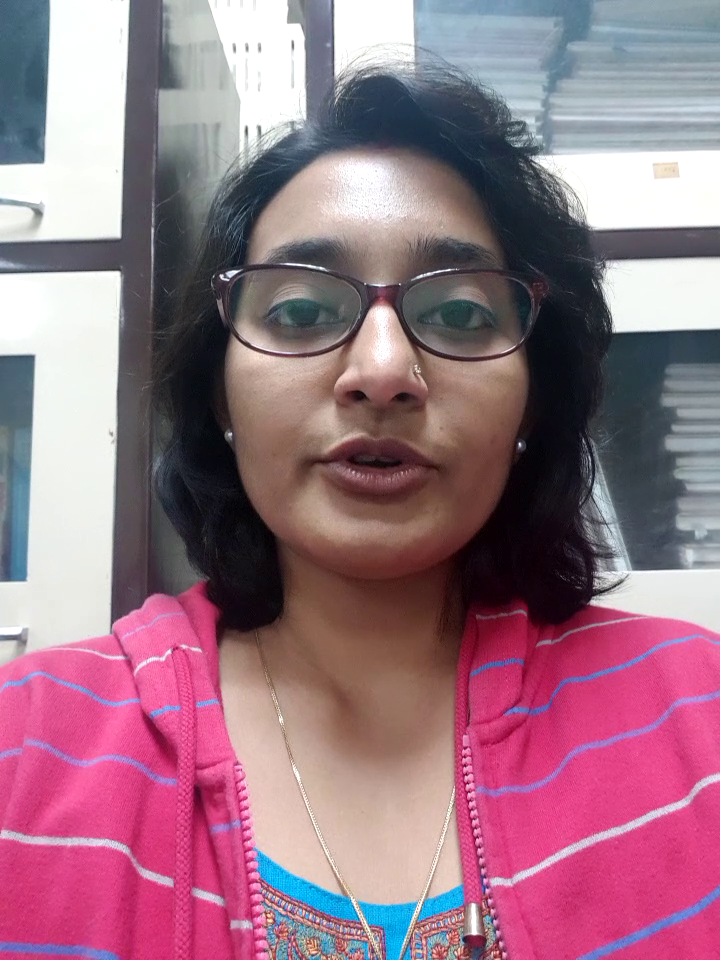}
    \\
    \includegraphics[width=.3\columnwidth]{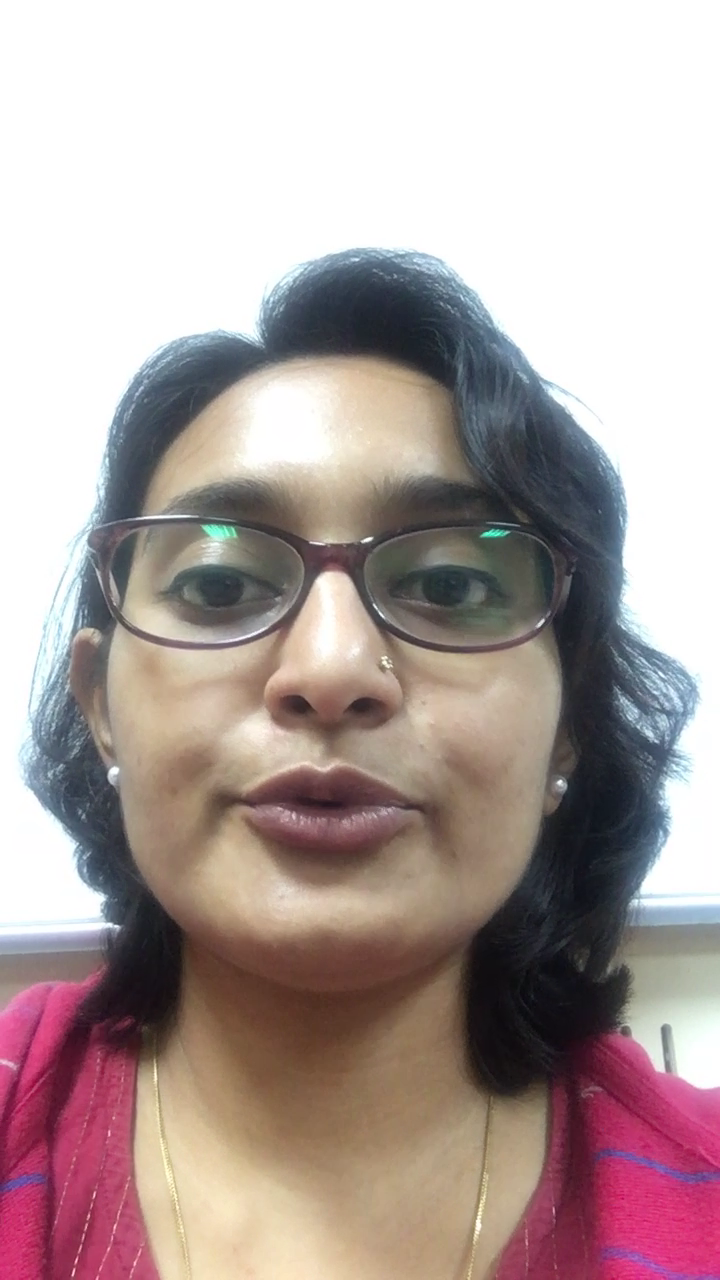}
    \includegraphics[width=.3\columnwidth]{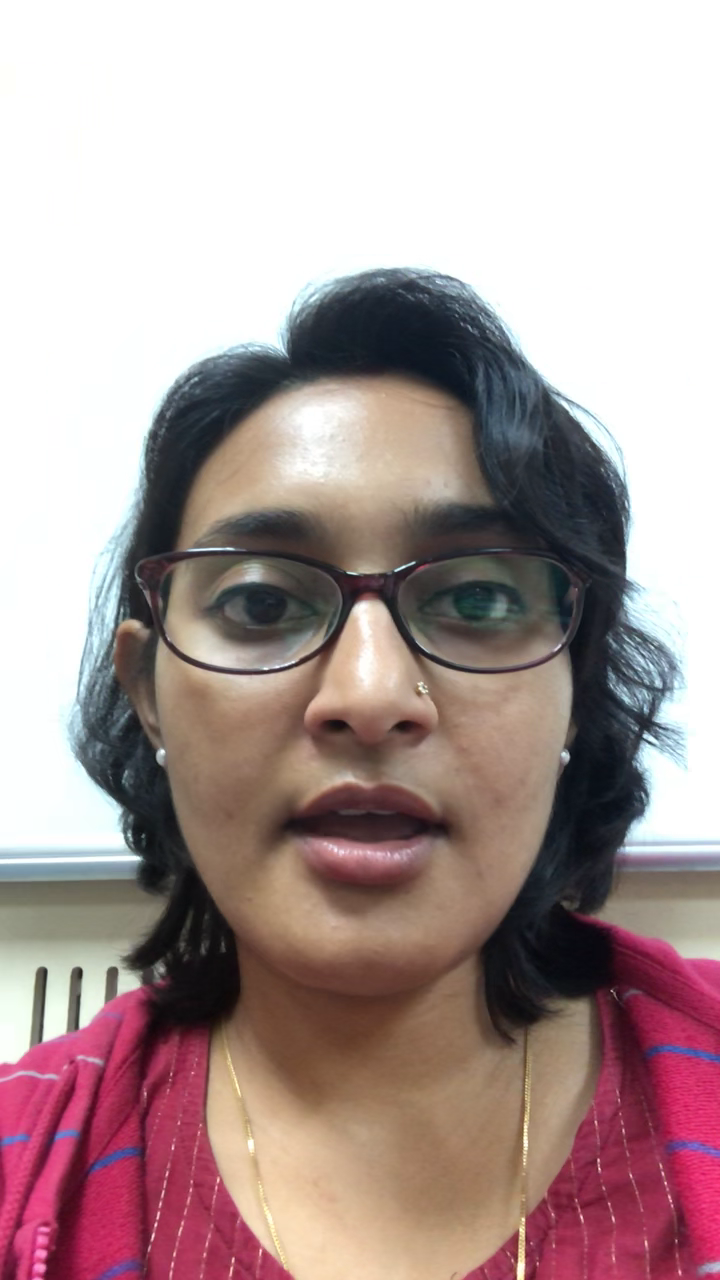}
    \includegraphics[width=.3\columnwidth]{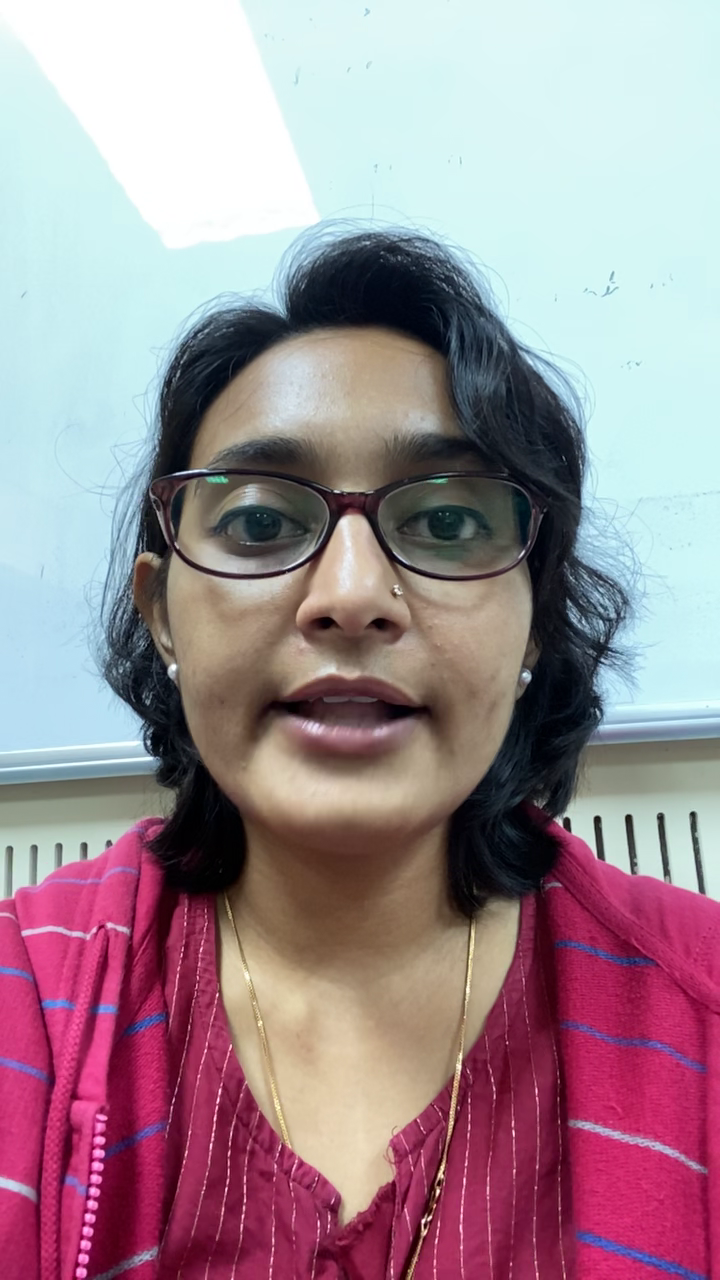}
    \includegraphics[width=.3\columnwidth]{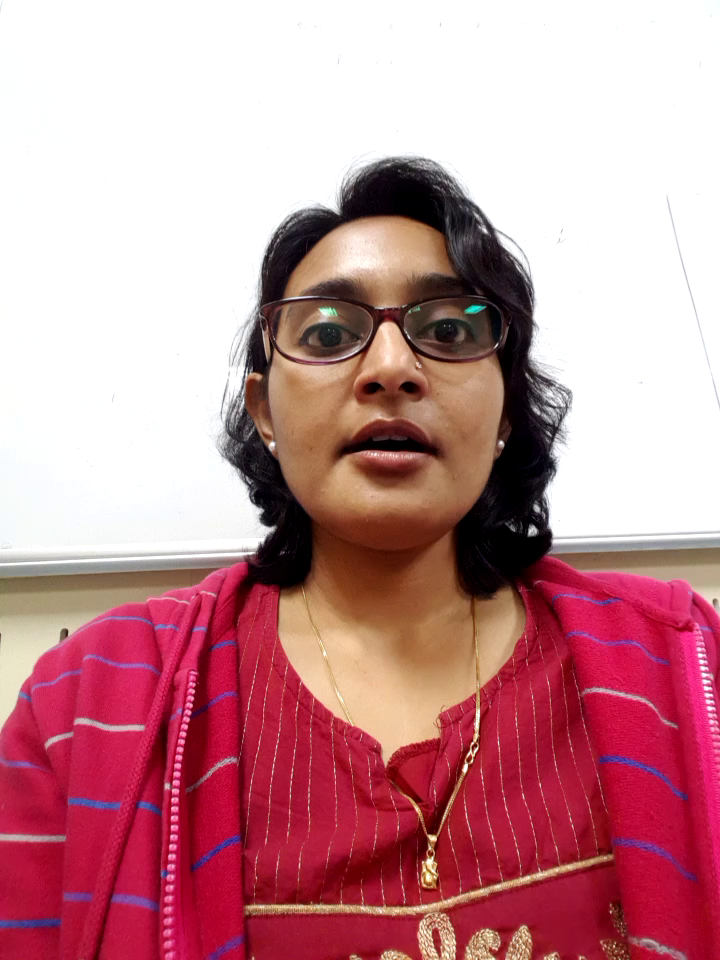}
    \includegraphics[width=.3\columnwidth]{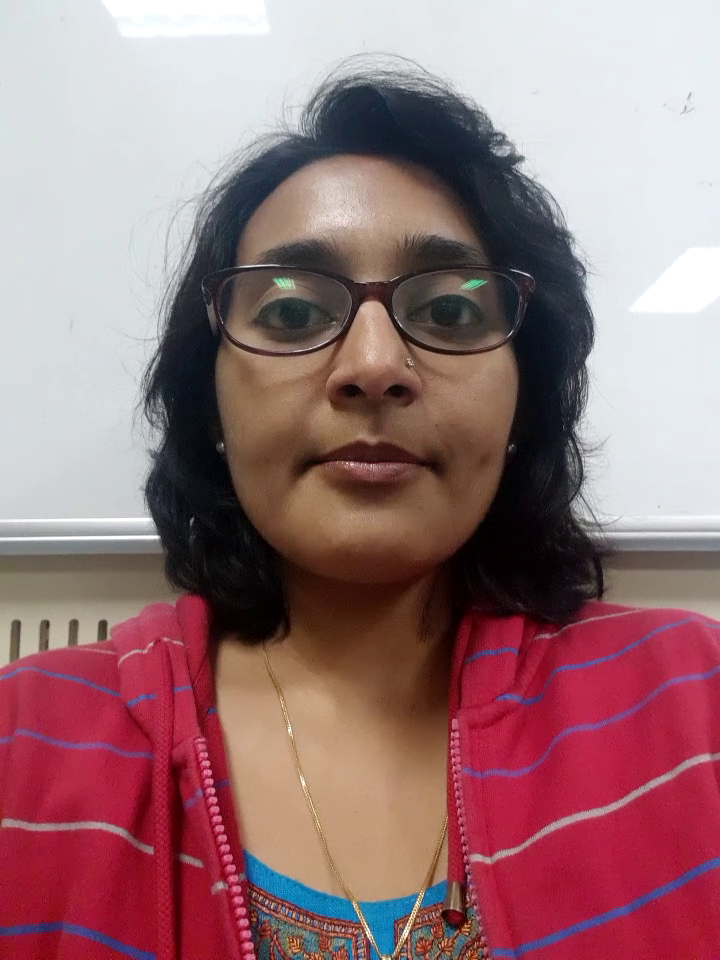}
    \\
    \includegraphics[width=.3\columnwidth]{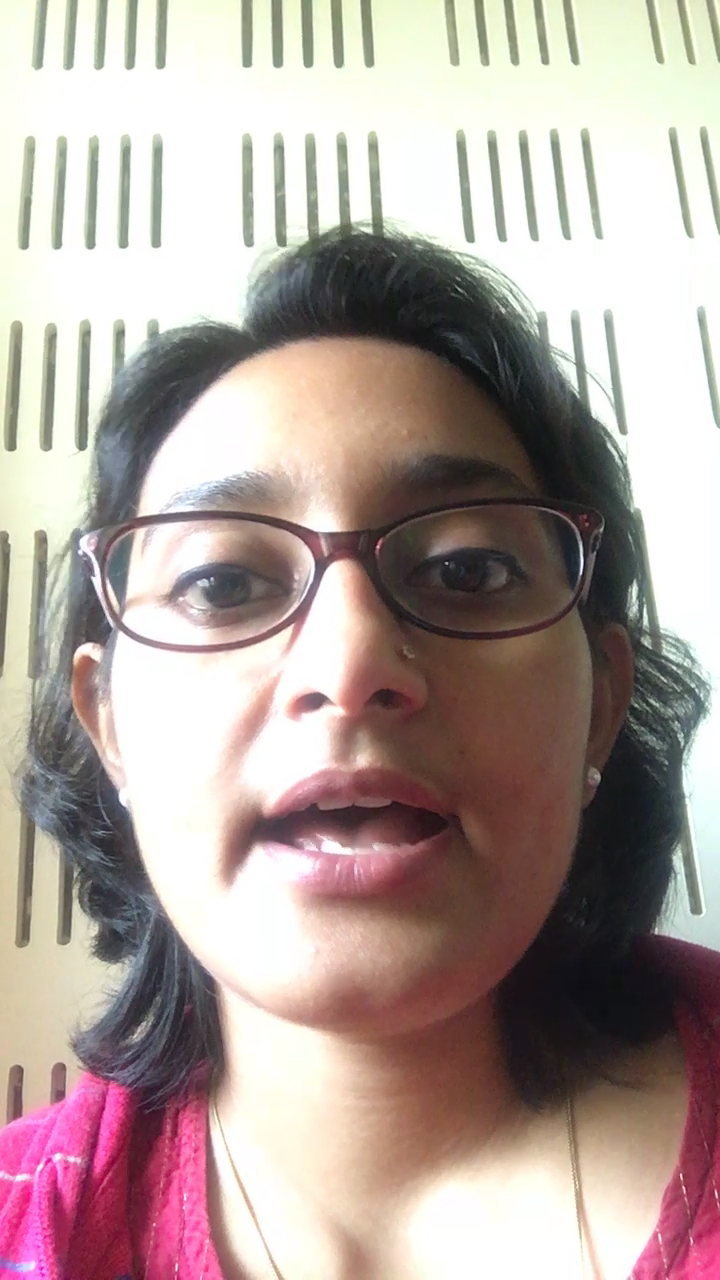}
    \includegraphics[width=.3\columnwidth]{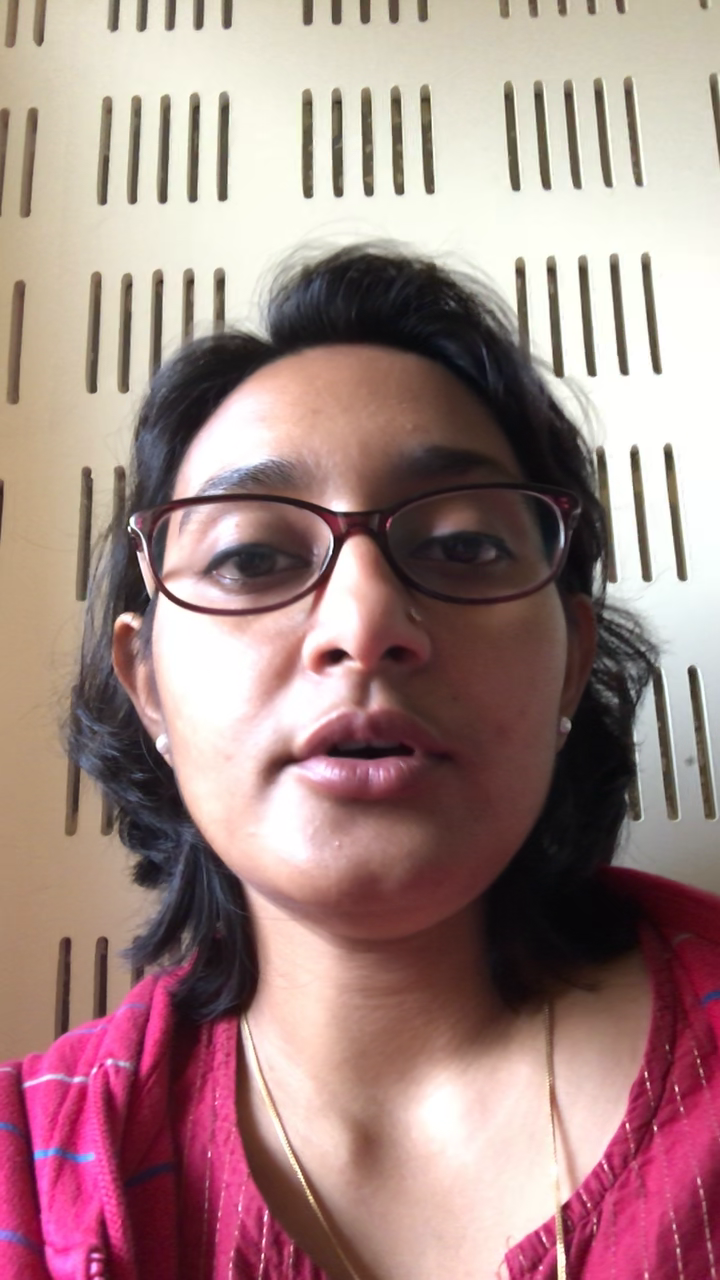}
    \includegraphics[width=.3\columnwidth]{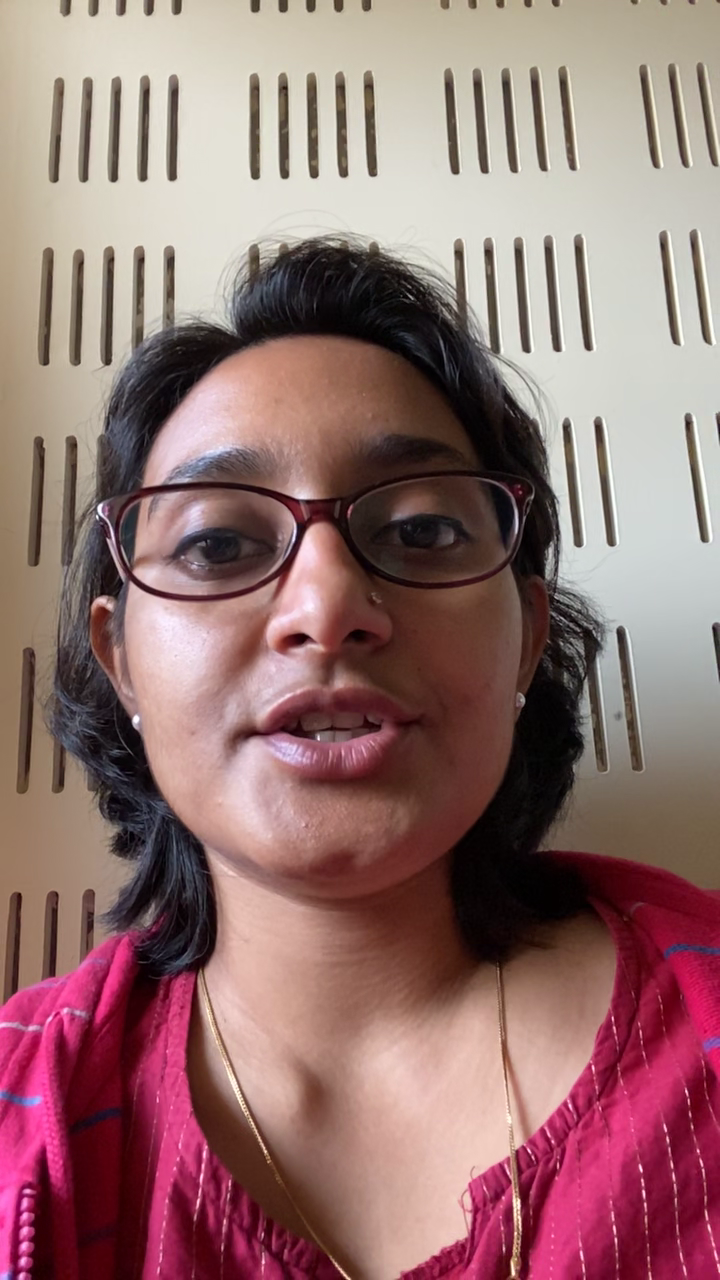}
    \includegraphics[width=.3\columnwidth]{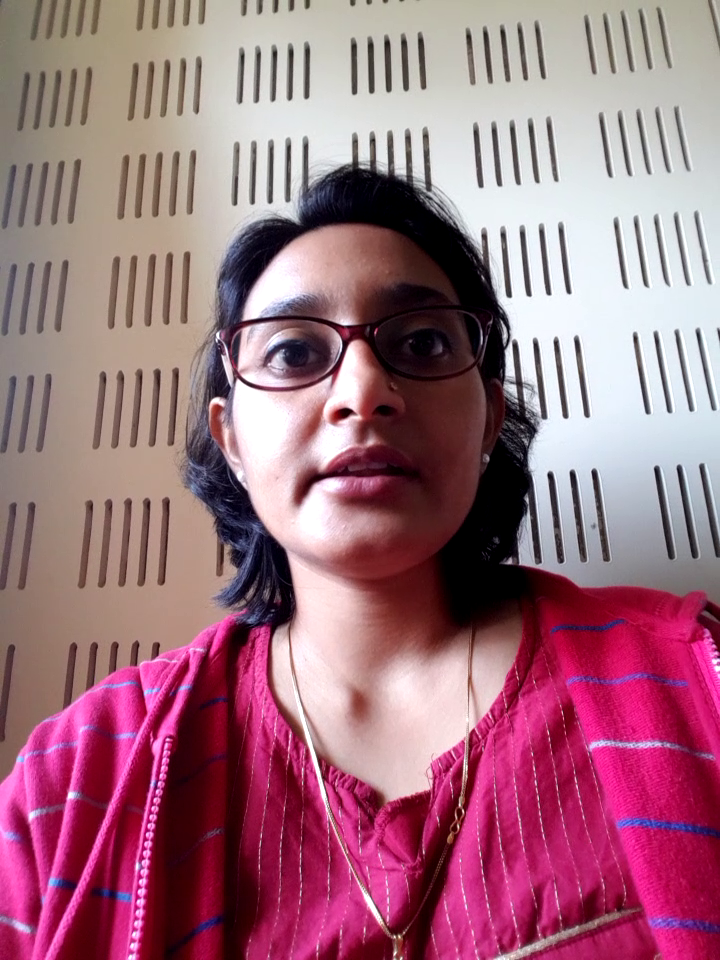}
    \includegraphics[width=.3\columnwidth]{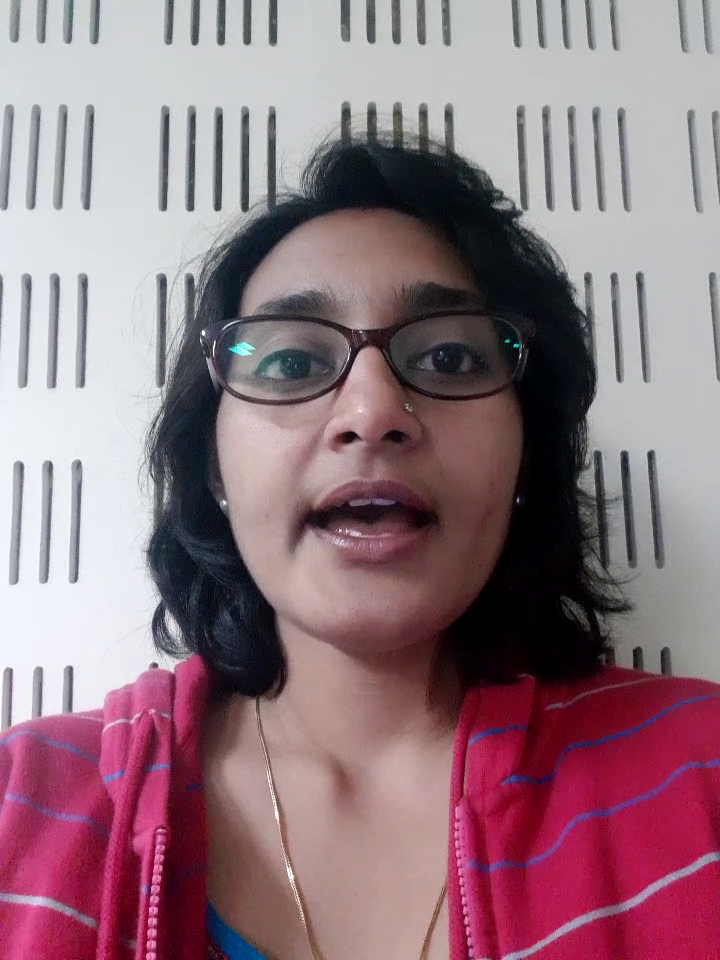}    \caption{Audio-visual data samples (1 frame of a talking face). Left to Right: iPhone 6s, iPhone 10, iPhone 11, Samsung S7 and Samsung S8. Top row: Session 1, middle: Session2, bottom: Session3.}
    \label{fig:data_sample_video}
\end{figure*}

\begin{figure*}
    \centering
    \includegraphics[width=.35\columnwidth]{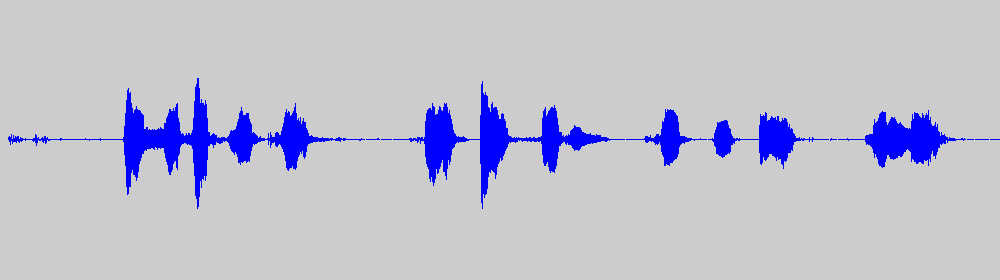}
    \includegraphics[width=.35\columnwidth]{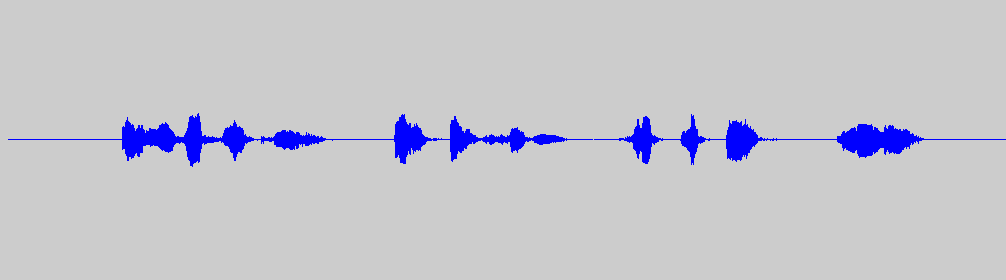}
    \includegraphics[width=.35\columnwidth]{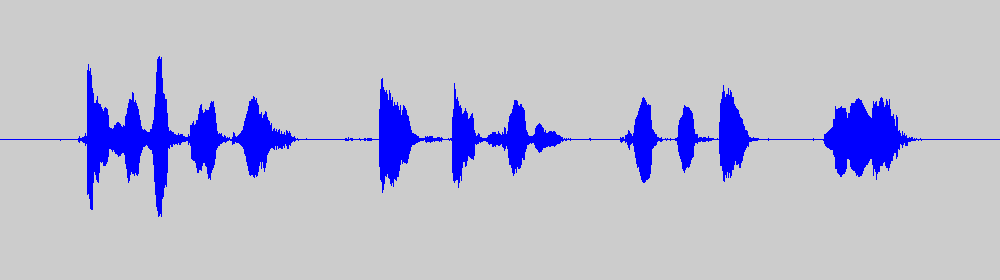}
    \includegraphics[width=.37\columnwidth]{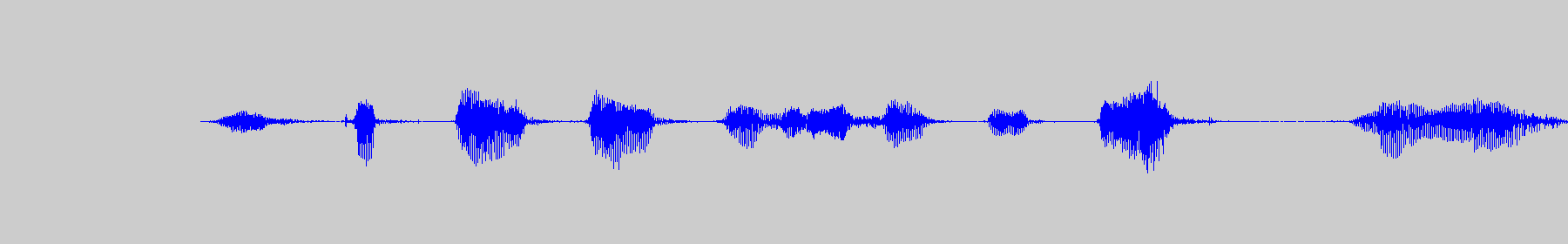}
    \includegraphics[width=.37\columnwidth]{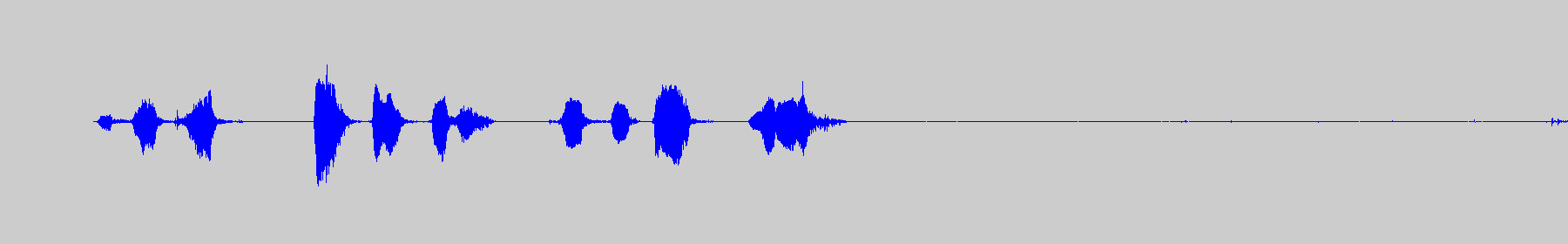}
    \\
    \includegraphics[width=.35\columnwidth]{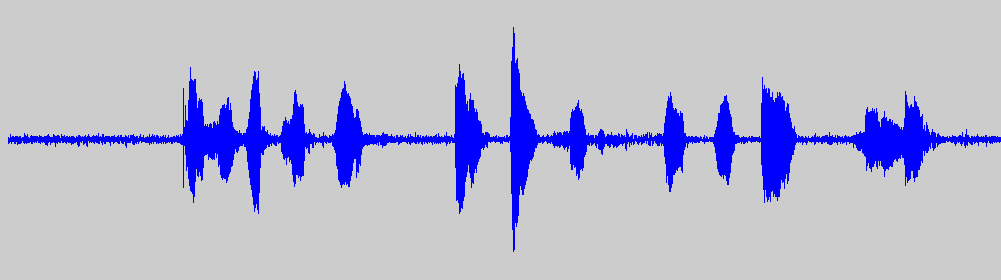}
    \includegraphics[width=.35\columnwidth]{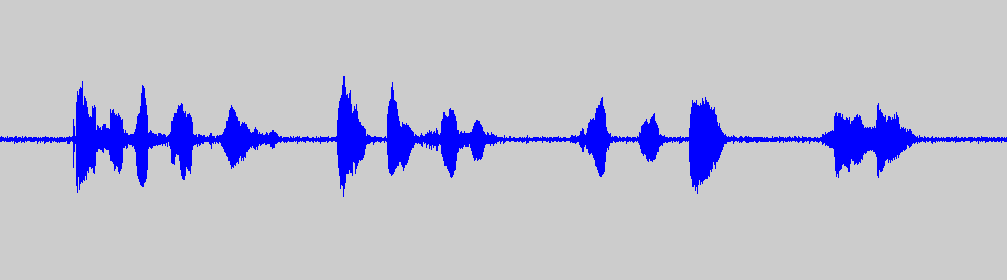}
    \includegraphics[width=.35\columnwidth]{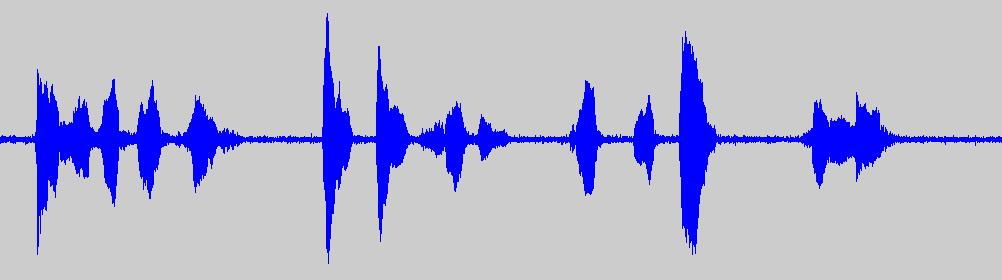}
    \includegraphics[width=.37\columnwidth]{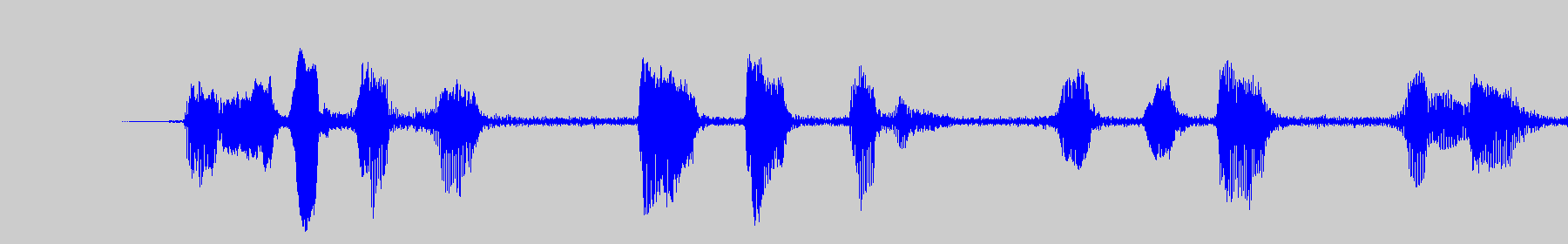}
    \includegraphics[width=.37\columnwidth]{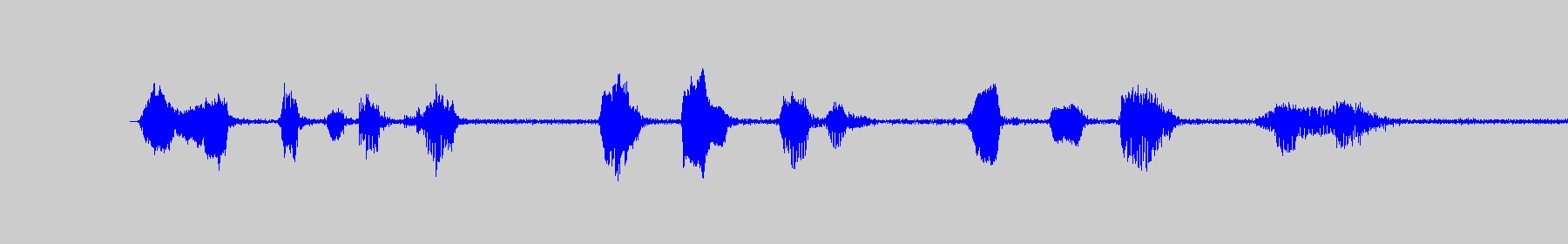}
    \\
    \includegraphics[width=.35\columnwidth]{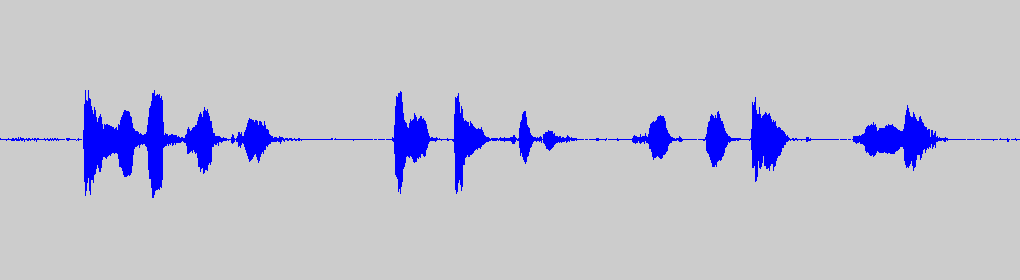}
    \includegraphics[width=.35\columnwidth]{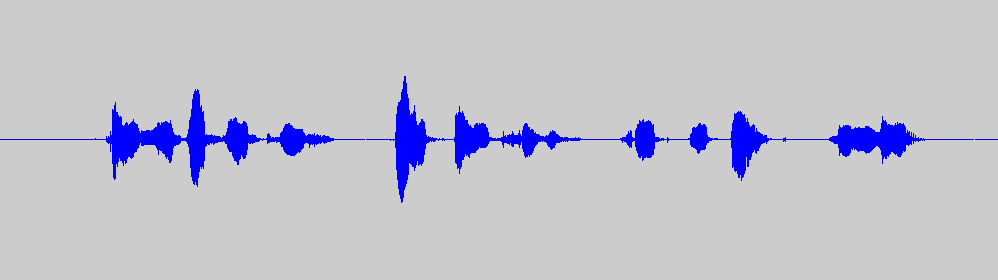}
    \includegraphics[width=.35\columnwidth]{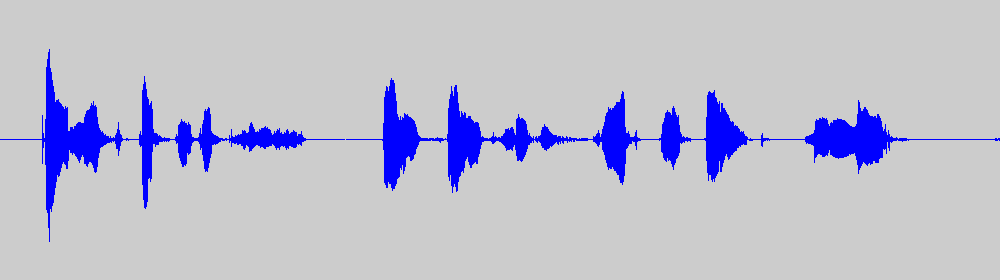}
    \includegraphics[width=.37\columnwidth]{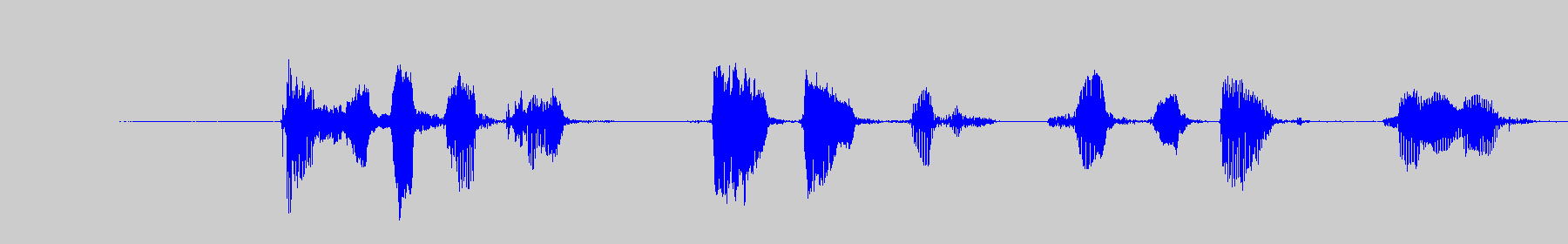}
    \includegraphics[width=.37\columnwidth]{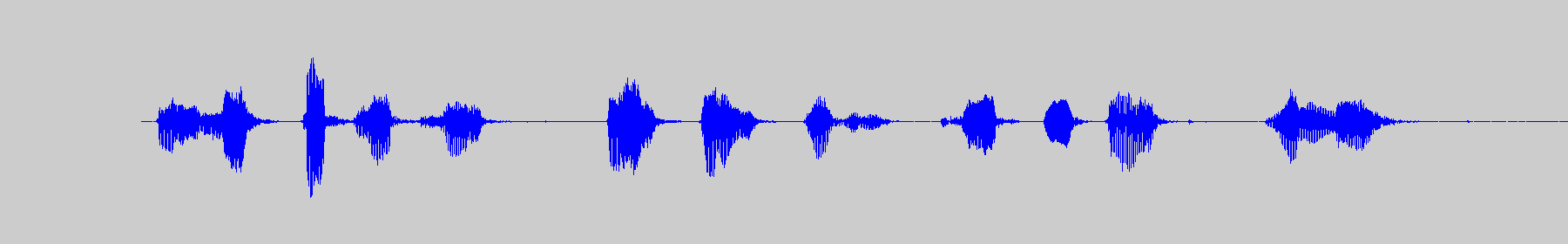}    
    \caption{Audio data sample for speaker recognition. Left to Right: iPhone 6s, iPhone 10, iPhone 11, Samsung S7 and Samsung S8. Top row: Session 1, middle: Session 2, bottom: Session 3.}
    \label{fig:data_sample_audio}
\end{figure*}

In data acquisition, we have used five smartphone devices, namely iPhone 11, iPhone10, iPhone 6s, Samsung S7 and Samsung S8. The data capturing is a self-assisted process where the speaker handles the mobile device and records the biometric data. For the process of data capturing, a mobile application has been used in both iOS and Android devices. The application provides a simple interface that assists the speaker to provide audio-visual data, as shown in Figure \ref{fig:app_mavsd}. A pre-defined text appears on the screen for a limited time for each sample. The speaker reads the text while the data is being recorded. 

\begin{figure*}[tbh]
    \centering
    \includegraphics[width=.35\columnwidth]{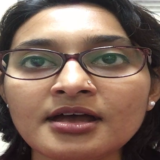}
    \includegraphics[width=.35\columnwidth]{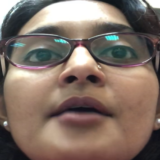}
    \includegraphics[width=.35\columnwidth]{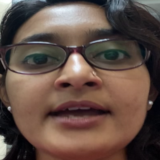}
    \includegraphics[width=.35\columnwidth]{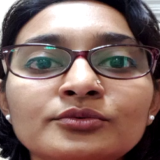}
    \includegraphics[width=.35\columnwidth]{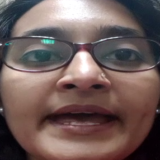}
    \\
    \includegraphics[width=.35\columnwidth]{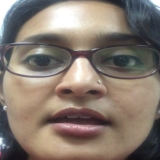}
    \includegraphics[width=.35\columnwidth]{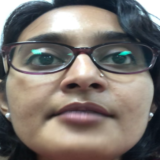}
    \includegraphics[width=.35\columnwidth]{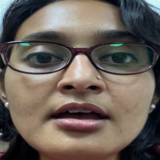}
    \includegraphics[width=.35\columnwidth]{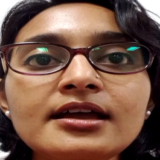}
    \includegraphics[width=.35\columnwidth]{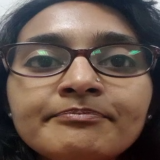}
    \\
    \includegraphics[width=.35\columnwidth]{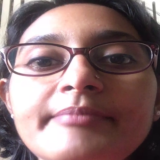}
    \includegraphics[width=.35\columnwidth]{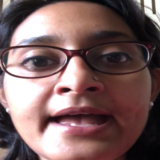}
    \includegraphics[width=.35\columnwidth]{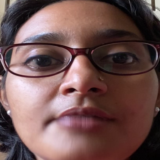}
    \includegraphics[width=.35\columnwidth]{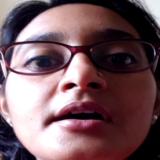}
    \includegraphics[width=.35\columnwidth]{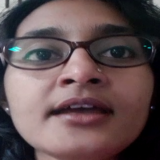}    
    \caption{Detected face using MTCNN for face recognition. Left to Right: iPhone 6s, iPhone 10, iPhone 11, Samsung S7 and Samsung S8. Top row: Session 1, middle: Session2, bottom: Session3.}
    \label{fig:data_sample_face}
\end{figure*}

\subsection{Participant details}
We have obtained 70 male and 33 female participants for the data collection. The average age of the participants is 27 years. All participants are of Indian origin with medium to expert range fluency in speaking the three languages (English, Hindi and Bengali). All participants are informed about the data acquisition protocol and are instructed to use the mobile application by self-assisting the data capture. Each session, the participant is given five mobile devices, one after the other, and audio-visual data of 6 sentences in three languages is recorded.

\subsection{Data details}

Each participant records six sentences in each language. Three of the sentences are the same for all subjects, and the other three sentences have a unique part for each subject. The six sentences in the English language are mentioned below, and the blank spaces are filled with unique fake text for each subject. Similarly, translated sentences for the other two languages are presented in their corresponding script.

\begin{enumerate}
    \item My full name is \texttt{fake name}.
    \item I live at the address \texttt{fake address}.
    \item I am working at IIT Kharagpur.
    \item My bank account number is \texttt{fake number}.
    \item The limit of my account is 10,000 rupees.
    \item The code for my bank is 9876543210.
\end{enumerate}

Data is captured in three sessions with three different lighting and noise environments. In session1, there is no noise, and uniform lighting is used. This data can be used as clean data for enrollment purposes. Session2 has continuous controlled noise from a portable fan intentionally put near the data capturing process and different lighting than session1 but with uniform illuminance. Session3 has uncontrolled noise from natural background and nonuniform lighting where certain parts of the participant's face are dark. The order of sentences, languages, and mobile devices used during data capture is kept the same for all the sessions. The sample video data can be seen in Figure \ref{fig:data_sample_video} (one frame per session, the device is presented for convenience). The waveform of audio samples is presented in Figure \ref{fig:data_sample_audio}. In Figure \ref{fig:data_sample_face}, the segmented face images (using MTCNN, see Section \ref{sec:mtcnn}) of each session and device are presented. 

\subsection{Presentation Attacks} \label{sec:pas}

We have created two types of presentation attacks: replay attacks and synthesized attacks. 

\subsubsection{Replay Attacks}

\begin{figure}[tbh]
    \centering
    \includegraphics[width=0.35\columnwidth]{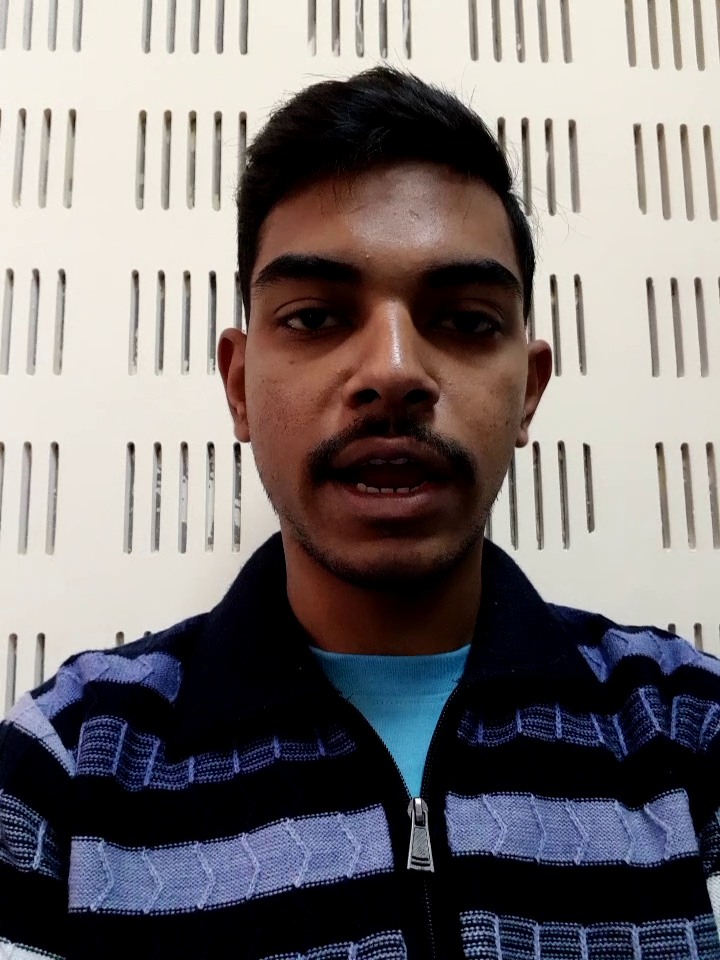}
    \includegraphics[width=0.35\columnwidth]{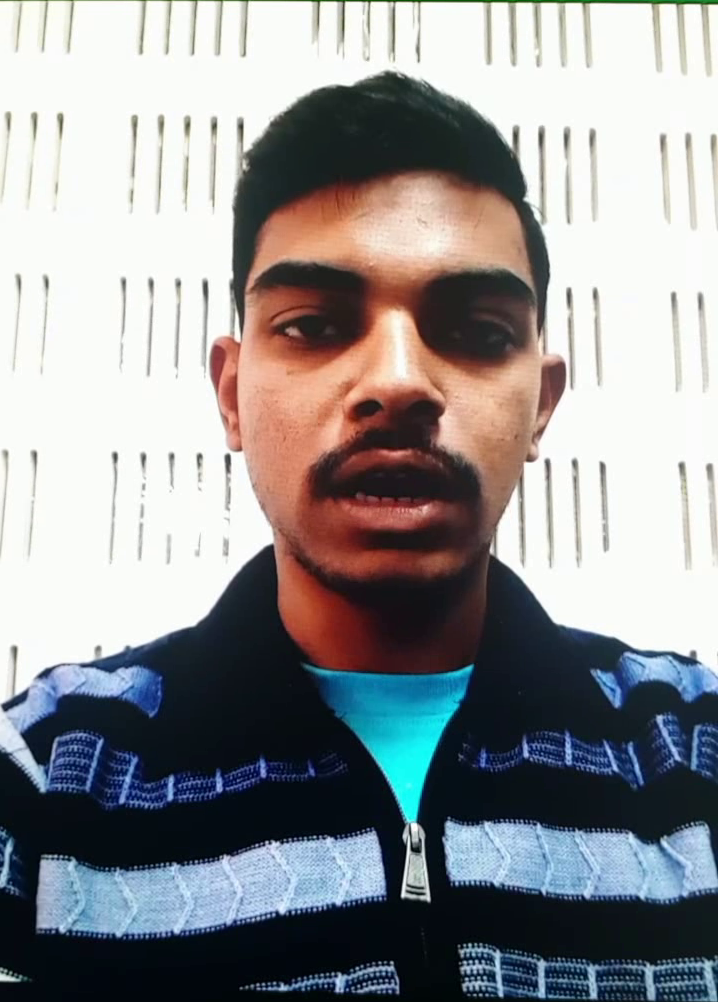}\\
    \includegraphics[width=0.35\columnwidth]{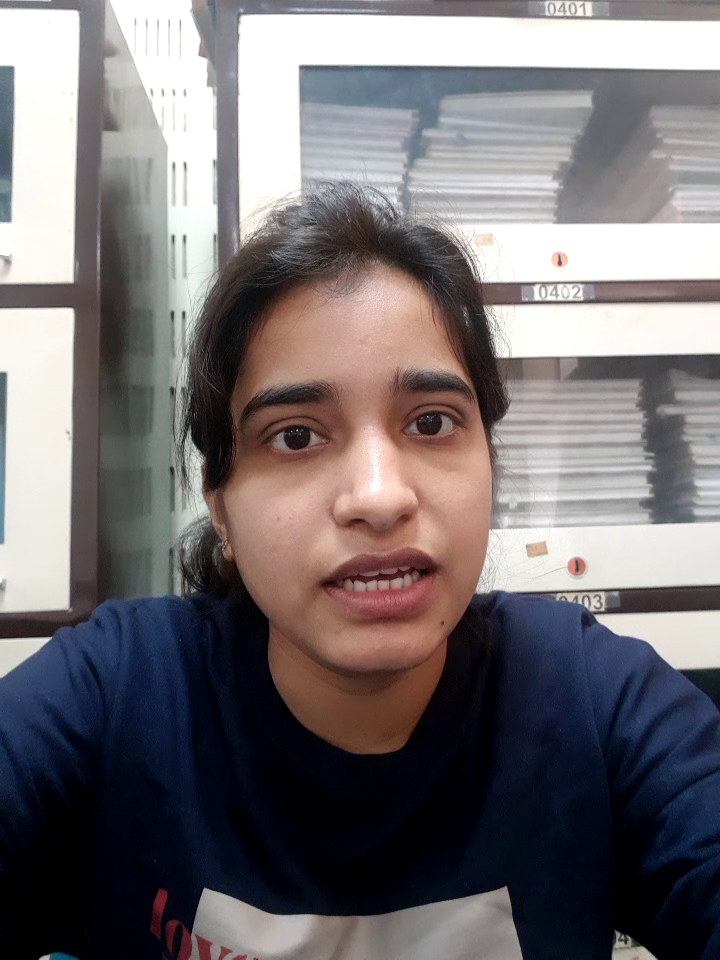}
    \includegraphics[width=0.35\columnwidth]{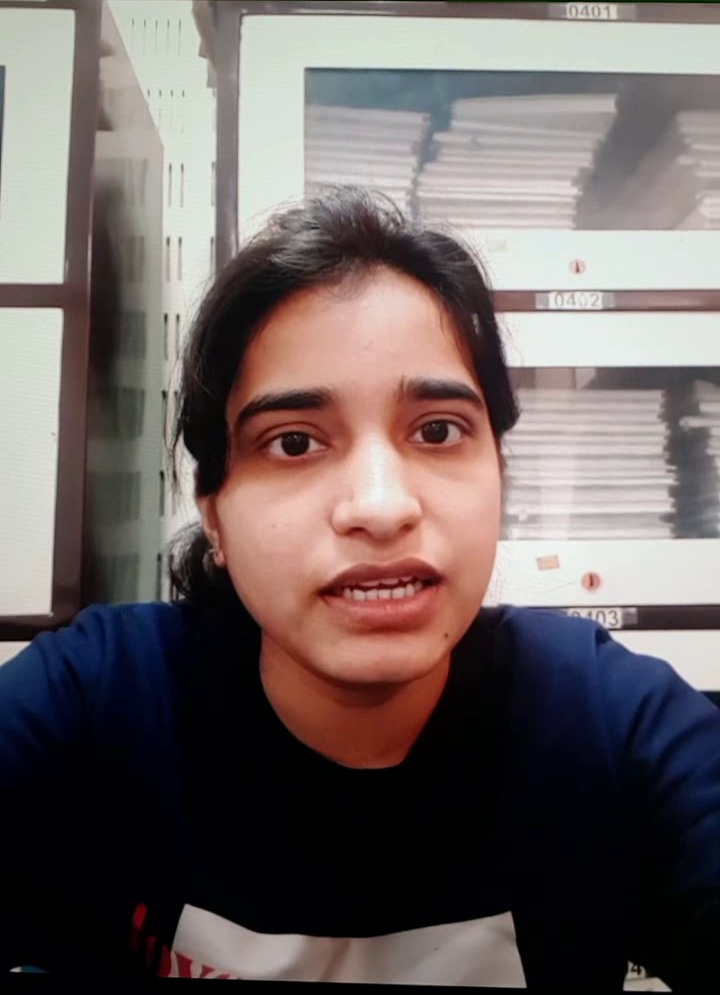}
    \caption{Replay attack data sample. Left: Bona fide, right: Replay attack.}
    \label{fig:face_replay}
\end{figure}

\begin{figure}[tbh]
    \centering
    \includegraphics[width=\columnwidth]{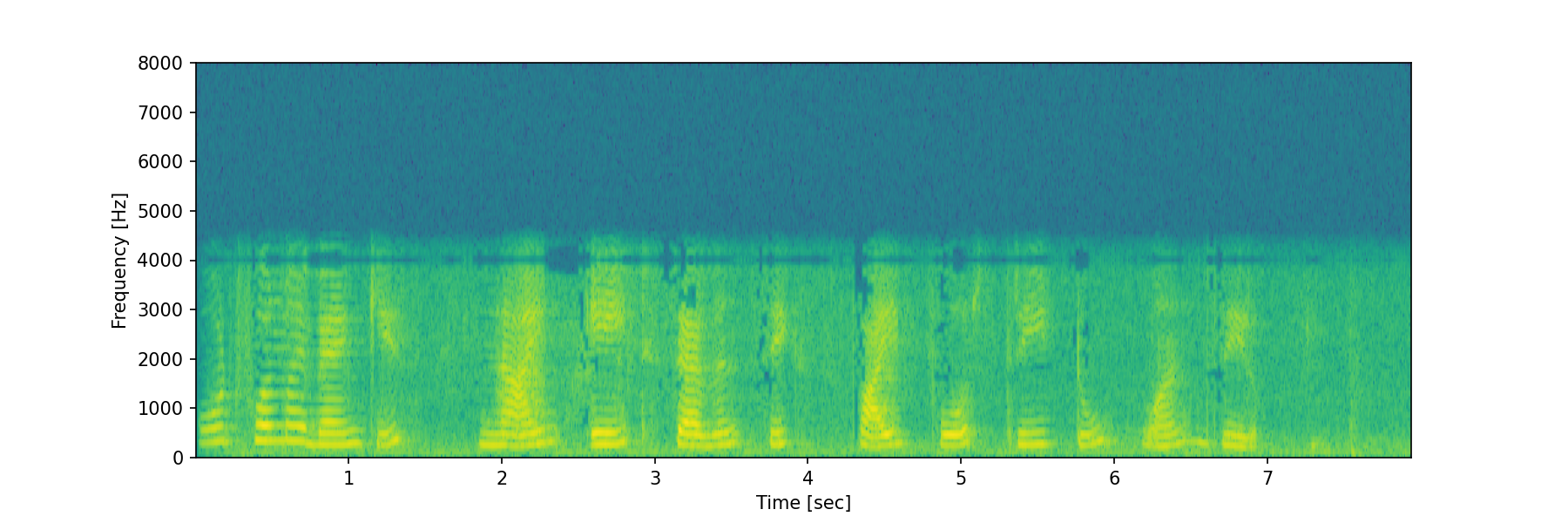}\\
    \includegraphics[width=\columnwidth]{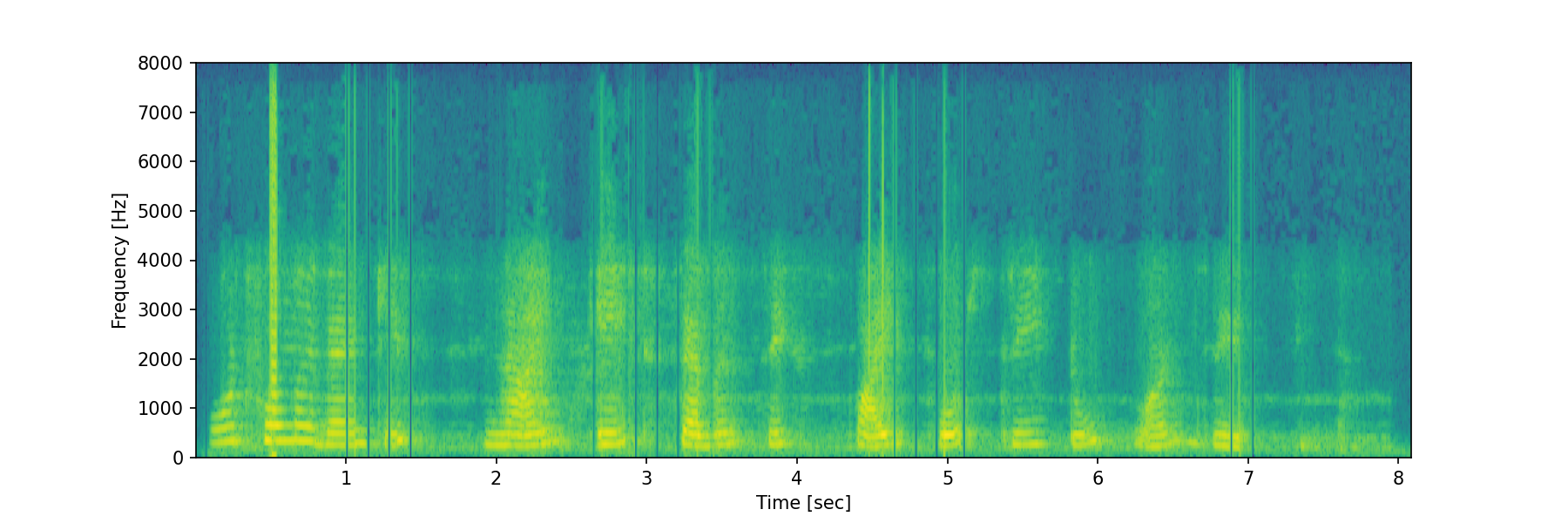}
    \caption{Spectrograms of bona fide and corresponding replay attack audio. Top: Bona fide, bottom: Replay attack.}
    \label{fig:spectrogram_replay}
\end{figure}

The replay attacks are created by synchronized capture of audio-visual playback using Dell office monitor and Logitech speakers recorded on Samsung S8 phone. Figure \ref{fig:face_replay} show the replay attacks samples created in this work. The spectrograms of audio replay attacks are presented in figure \ref{fig:spectrogram_replay}.

\subsubsection{Synthesized attacks} \label{sec:synth_attacks}

Deep learning has been successfully applied to solve complex problems ranging from big data analysis to computer vision tasks and human level control. Advanced deep learning concepts have also been used to create threats to privacy, democracy and national security. One such deep-learning based application that loomed recently is "deepfake" (derived from 'deep learning' and 'fake'). For creating synthesized attacks, we have used deepfake approaches in this work. One of the approaches for creating face deepfakes is a technique where the face image of a source person is superimposed onto a target person to create a video/image of the target person. In this direction, the face-swapping model is proposed by Nirkin \textit{et al.} \cite{nirkin2019fsgan} where swapping of face images are done in three stages. Reenactment and face segmentation is carried out in the first stage, followed by in-painting and blending. Reenactment, face transfer, or puppeteering uses facial expressions and assists in transforming the face in one video to guide the motions and deformations of the face appearing in another video or image. Face segmentation is performed using U-Net \cite{ronneberger2015u} and reenactment is performed using generative model named pix2pixHD \cite{wang2018high}. In the second step, the occluded regions of the source face are mitigated using the same in-painting generator \cite{wang2018high}. In the last step, a Gaussian Poisson Generative Adversarial Network (GP-GAN) \cite{wu2019gp} is used for high-resolution image blending for combining the gradient and colour information. 

In our work, we have utilized FSGAN for swapping similar faces \footnote{FSGAN: \url{https://github.com/YuvalNirkin/fsgan}}. The face-swapping approach preserves the context of the target video by digitally overlaying the source's face landmarks. Therefore, the target video contains the key biometric characteristics of the source subject, which can efficiently be used as a presentation attack for the source's identity. Multiple deepfake datasets in the literature \cite{yang2019exposing, korshunov2018deepfakes, rossler2019faceforensics++, dolhansky2019deepfake} used a manual selection of faces for swapping. However, we have employed an automatic way to find a pair of similar faces in this work. We used cosine similarity of ArcFace embeddings to find a similar face for each of the male and female subjects (more on ArcFace in section \ref{sec:arcface}). We have generated 97 face swapped videos for sentence 6 of bona fide data from session1 data of the Samsung S8 device.


\begin{figure}[tbh]
    \centering
    \includegraphics[width=0.3\columnwidth]{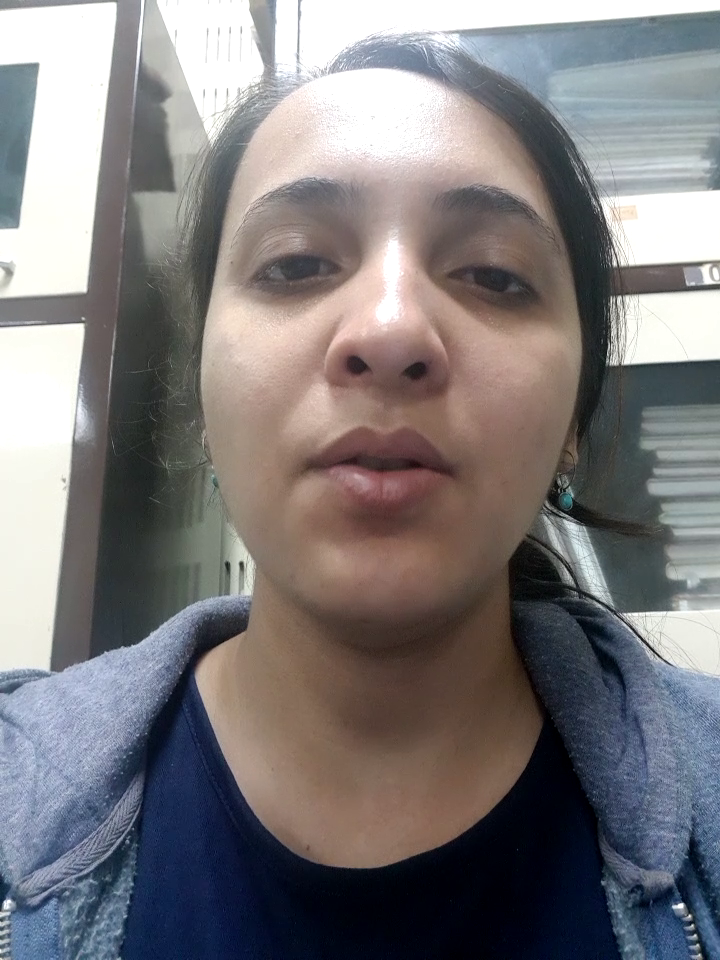}
    \includegraphics[width=0.3\columnwidth]{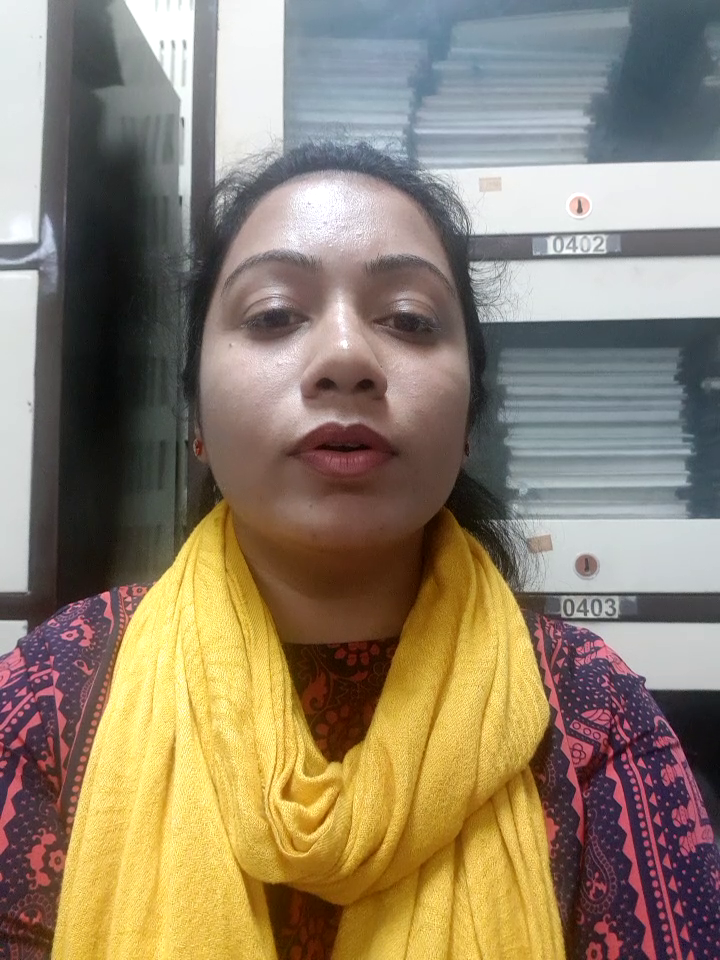}
    \includegraphics[width=0.3\columnwidth]{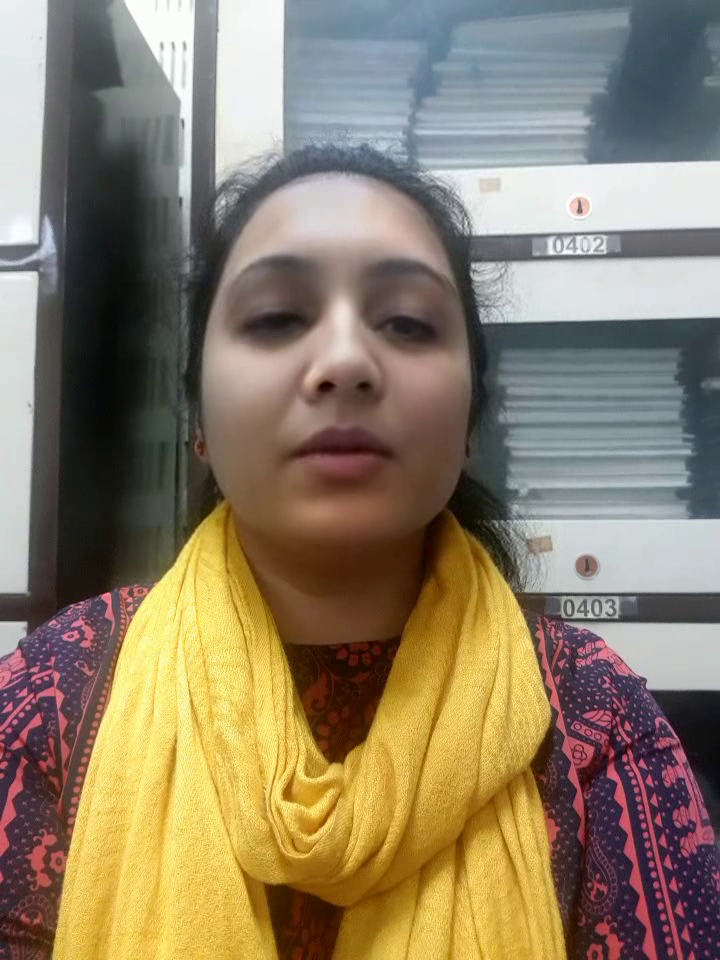}\\
    \includegraphics[width=0.3\columnwidth]{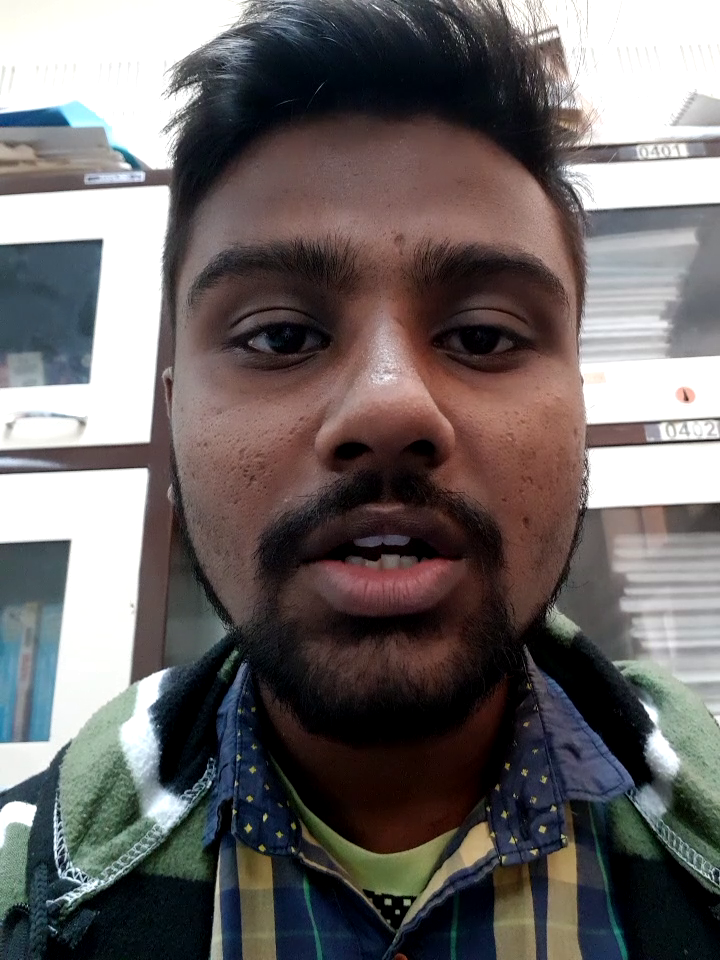}
    \includegraphics[width=0.3\columnwidth]{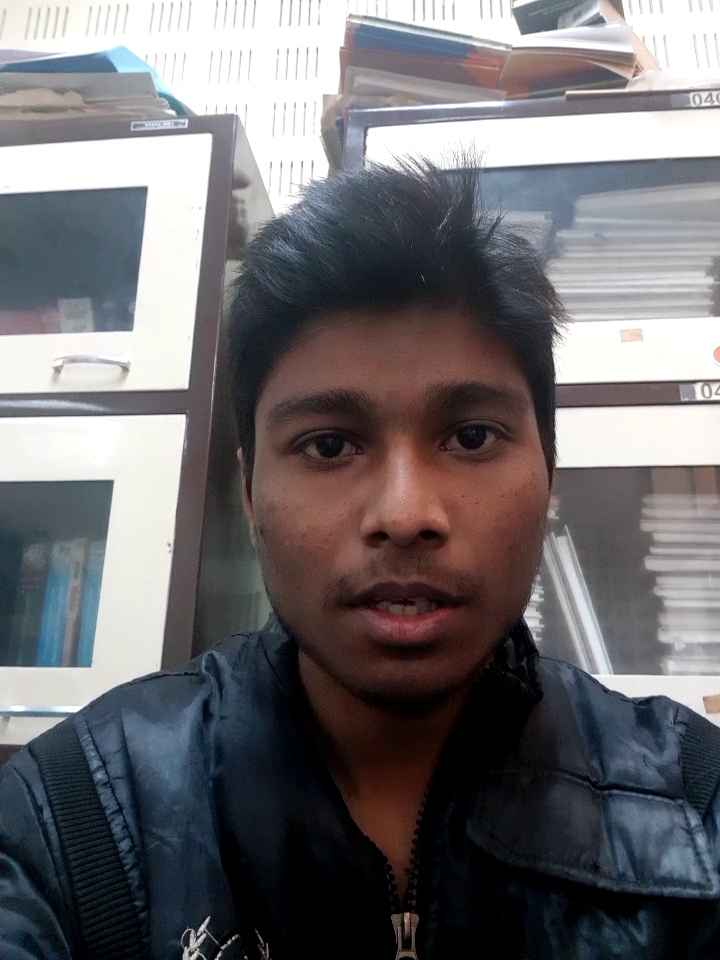}
    \includegraphics[width=0.3\columnwidth]{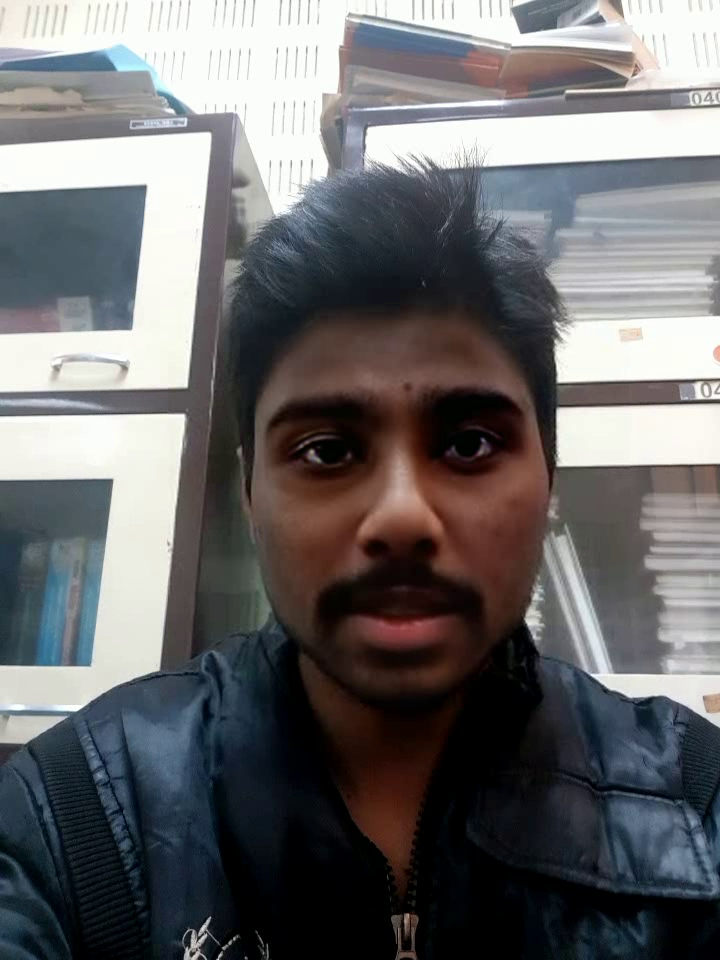}
    \caption{Face swap using FSGAN. Left: Source face, middle: Target face, right: Swapped face.}
    \label{fig:fsgan_sample}
\end{figure}

\begin{figure}[tbh]
    \centering
    \includegraphics[width=\columnwidth]{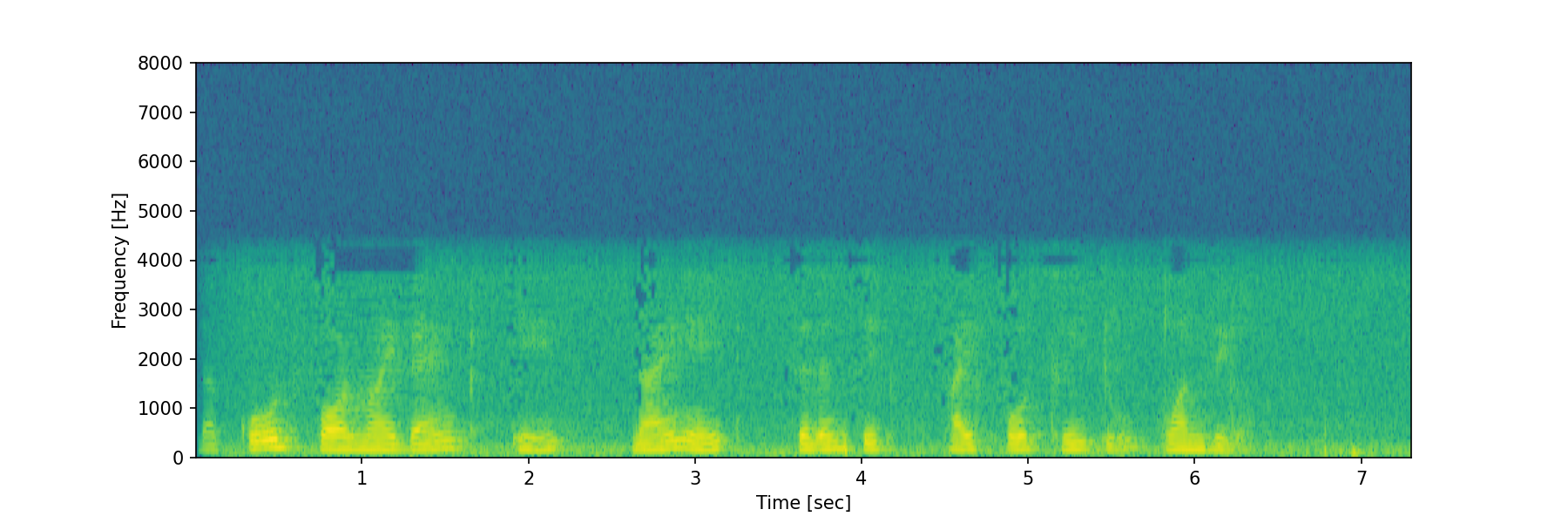}\\
    \includegraphics[width=\columnwidth]{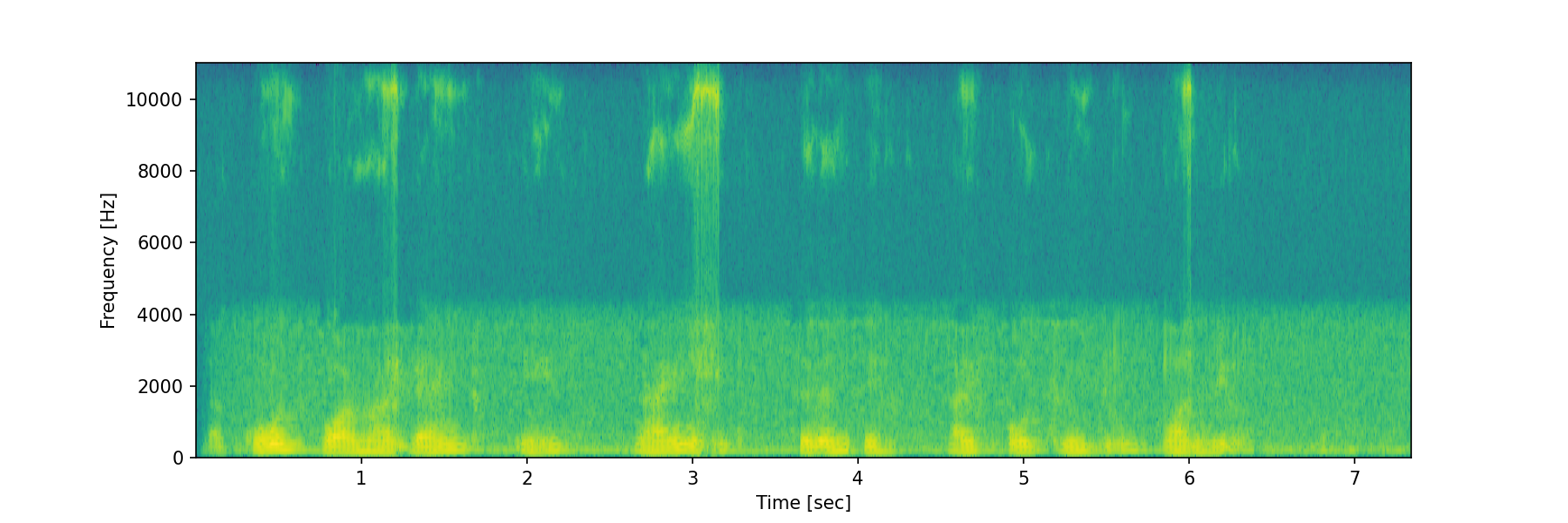}
    \caption{Spectrograms of bonafide and corresponding wavenet-vocoder synthesized audio. Top: Bona fide, bottom: Synthesized audio.}
    \label{fig:spectrogram_wavenet}
\end{figure}

WaveNet vocoder is used to generate high-quality raw speech samples conditioned on acoustic features \cite{oord2016wavenet}. The WavNet-based vocoder is popularly used in ASVSpoof 2019 challenge to create logical access presentation attacks \cite{todisco2019asvspoof}. In our work, we have used MFCC features as acoustic features in synthesizing 16-bit raw audio. We have adapted the implementation of WaveNet vocoder form the github\footnote{WaveNet Vocoder: \url{https://github.com/r9y9/wavenet_vocoder}} and pre-trained models from LJSpeech \cite{ljspeech17}. The figures \ref{fig:fsgan_sample} and \ref{fig:spectrogram_wavenet} show the images samples and spectrograms of synthesized attacks respectively.

\section{Performance Evaluation Protocols} \label{sec:perform_eval}

The dataset is benchmarked with various face recognition, speaker verification and presentation attack detection methods. In this section, we explain briefly the baseline biometric systems employed along with evaluation metrics. 

\subsection{Automatic speaker Verification}

\subsubsection{I-vector based speaker Verification}

The I-vector based ASV method is a Joint Factor Analysis (JFA) approach proposed in \cite{dehak2011front}. It models the channel effects and also speaker voice characteristics. The speech sample is represented as a low-dimensional super vector called i-vector. The i-vector represents the total factor in a speech utterance, including channel compensation which is carried out in a low-dimensional total variability space. 

\subsubsection{X-vector based speaker Verification}

The deep neural networks (DNN) and end-to-end speaker verification approaches are state-of-the-art research methods that overcome handcrafted methods' drawbacks. The x-vector based speaker verification is a recent approach showing promising results in automatic speaker verification \cite{xvector2018snyder}. This method uses deep neural network (DNN) embeddings as features. The variable-length speech utterances are mapped to a fixed low-dimensional embedding (called x-vectors), and a deep network is trained to differentiate speakers. The training process requires a large amount of training data. Therefore, data augmentation is used along with added noise and reverberation to increase training data size. 
The implementations in Kaldi are employed in our work, and the pre-trained Universal Background Models, i-vector extractor and x-vector extractor are adapted to our experiments \footnote{Kaldi GitHub: \url{https://github.com/kaldi-asr/kaldi}}. Probabilistic linear discriminant analysis (PLDA) \cite{prince2011probabilistic} is used as a classifier for the i-vectors and x-vectors of enrollment and test samples. The log-likelihood score is computed between the enrolled and test speech sample pair.  

\subsubsection{Dilated residual network (DltResNet)}

Extended ResNet implementation from \cite{le2018robust} named dilated residual network (DltResNet) is used as the third speaker verification methods. The implementation is publicly available\footnote{DltResNet: \url{https://www.idiap.ch/software/bob/docs/bob/bob.learn.pytorch/v0.0.4/guide_audio_extractor.html}}. The DltResNet model is one of the state-of-the-art systems on the Voxceleb1 database evaluations achieving 4.8\% EER on the dataset. The Euclidean distance between the DltResNet features is used for obtaining scores between enrolled and test samples.

\subsection{Face recognition}

\subsubsection{Face Detection} \label{sec:mtcnn}

Face detection is performed as a prepossessing step on the video frames to detect and crop the face image. We have employed multitask cascaded convolutional networks (MTCNN) approach from Zhang \textit{et al.} \cite{zhang2016joint} for efficient face detection. The face recognition and face PAD methods used in this work used segmented face images.

\subsubsection{Local Binary Patterns (LBP)}

Local Binary Patterns (LBP) are a textual operator that labels the pixels in a face image according to neighbouring pixels' values and assigns a binary number. LBP for an image is calculated by assigning 0 or 1 to the pixel depending on the neighbour's pixel having high or low value. The resultant binary test is stored in an 8-bit array and later converted to decimal. This thresholding process, accumulating binary strings, and storing the decimal value is repeated for every pixel in the input image. Further, the LBP histogram is computed over the LBP output array. For a block, one of the $2^8=256$ possible patterns is possible. The advantage of LBP features is high discriminative power, computational simplicity, and invariance to grey-scale changes. LBPs have shown a prominent advantage in face recognition approaches. We used LBP histograms as features for face images and cosine distance to compute the score between the enrolled and test samples.

\subsubsection{FaceNet face embeddings}

The deep learning approaches have evolved into image processing and pattern recognition applications. In face recognition methods, FaceNet embeddings displayed an excellent image representation for facial features \cite{schroff2015facenet}. This is a deep face recognition approach that adapted the ideas from \cite{parkhi2015deep}. In this work, we have used the pretrained model on the VGGFace2 dataset using Inception ResNet v1. This model displayed an accuracy of 99.65\% on the Labeled Faces in the Wild (LFW) dataset \cite{huang2008labeled}. We have obtained FaceNet embeddings \footnote{FaceNet: \url{https://github.com/davidsandberg/facenet}} for face detected images in our dataset and used cosine distance between the samples to obtain the verification scores.

\subsubsection{ArcFace face descriptor} \label{sec:arcface}

ArcFace face features are proposed in \cite{deng2019arcface} for the large scale face recognition with enhanced discriminative power. ArcFace features emphasize the loss function in deep convolutional neural networks (DCNN) for clear geometric interpretation of face images. The proposed descriptor is evaluated over ten face recognition benchmarks, and results show consistent performance improvement. We have employed the ArcFace implementation provided in Github \footnote{ArcFace: \url{https://github.com/deepinsight/insightface}}. The training data contains cleaned MS1M, VGG2 and CASIA-Web face datasets. ArcFace face descriptors are computed over detected face images, and similar to other face recognition methods, we have used cosine distance as a classifier. 

In addition to the face recognition, we have used ArcFace face embeddings to obtain similarity scores between subjects in creating attacks in FSGAN face swapped videos (see section \ref{sec:synth_attacks}). 

\subsection{Presentation Attack Detection (PAD)} \label{sec:pad_methods}

\subsubsection{Voice PAD}

The PAD methods used to evaluate the attacks created using speech are chosen from the baseline methods in the ASVSpoof 2019 challenge \cite{todisco2019asvspoof}. The two baseline methods are available in ASVSpoof 2019 evaluation protocols. Features used in these two methods are based on cepstral coefficients in the front-end and Gaussian Mixture Models (GMM) in the back-end. Linear Frequency Cepstral Coefficients (LFCC) and Constant Q Cepstral Coefficients (CQCC) are two features used to represent speech samples.

The LFCC features are similar to the Mel-frequency cepstral coefficients (MFCCs), with filters placed linearly in the exact sizes. The initial approach of LFCCs is used for the detection of synthetic speech in \cite{sahidullah2015comparison}. In this work, we used LFCC features are extracted with a frame length of 25ms and a 20-channel linear filter bank. An LFCC feature comprises 19 cepstral coefficients, a zeroth coefficient, static, delta, and delta-delta coefficients. The CQCC features are extracted with the toolkit provided in ASVSpoof 2019. The maximum frequency is set to fs/2, where fs is the sampling frequency, and the minimum frequency is fixed at $fs/2/2^9 ~15Hz$ (where 9 is the number of octaves) \cite{todisco2016new}. The number of bins per octave is set to 96, and re-sampling is applied with a period of 16. The dimension of features is 29 coefficients along with zeroth, static, delta, and delta-delta coefficients.

The front-end provides the cepstral coefficients, which are used to train 2-class GMMs in the back-end. The training process is carried out on the bonafide and attack speech samples with 512-component GMM models. An expectation-maximization (EM) algorithm is employed in training with random initialization. For testing, the scores of samples are calculated from the log-likelihood ratio with the help of trained bona fide and the attack speech models.

\subsubsection{Face PAD}

The face recognition PAD methods are chosen from the baseline methods used in smartphone dataset evaluation in \cite{ramachandra2019smartphone}. The two best-performing methods from five baseline methods are taken for evaluation in this work. These methods utilize local binary patterns (LBP) \cite{chingovska2014biometrics} and color texture features \cite{boulkenafet2015face}. The support vector machines (SVM) are trained for different attacks and test for attack detection.

The LBP features are experiments for PAD in \cite{chingovska2014biometrics} for face attacks in a full biometrics verification system. In \cite{ramachandra2017presentation}, the LBP features displayed a consistent performance of detecting attacks in different protocols of smartphone biometric data. Similarly, the experiments using colour texture features \cite{boulkenafet2015face} resulted in the best-performing face PAD on smartphone face images. Therefore, we have included these methods in our evaluation of detection attacks.

\subsection{Performance Metrics}

The performance evaluation metrics from ISO/IEC \cite{ISO-IEC-19795-4-071128} are utilized in our experiments to present and compare the results of different methods.

\subsubsection{Verification Metrics}

\begin{itemize}
    \item False Match Rate (FMR) is the proportion of the completed biometric non-mated comparison trials that result in a false match. 
    \item False Non-Match Rate (FNMR) is the proportion of the completed biometric mated comparison trials that result in a false non-match. 
\end{itemize}

In addition to ISO/IEC metrics mentioned above, we have also presented an equal error rate (EER) to represent FMR and FNMR metrics in a single value. EER is the error rate at the point where FMR and FNMR are equal.    

\subsubsection{Presentation Attack Detection Metrics}

\begin{itemize}
    \item Impostor-Attack Presentation Match Rate (IAPMR) is the proportion of impostor attack samples (replay attacks) that are matched with bona fide samples. To compare ASV methods' performance, we have fixed FMR at 0.1\% and presented FNMR and IAPMR for zero-effort impostors and attacks, respectively.
    \item Attack Presentation Classification Error Rate (APCER) is the proportion of attack presentations that are incorrectly classified as bona fide presentations, and Bonafide Presentation Classification Error Rate (BPCER) is the ratio of bona fide presentations incorrectly classified as attacks. This work presents the BPCER\_5 and BPCER\_10 of PAD methods: the BPCER values at APCER are 5\% and 10\%, respectively.
\end{itemize}

\begin{table*}[!ht]
    \centering
    \caption{Inter-session speaker recognition evaluation (EER\%).}
    \label{tab:audio_session_ee}
    \begin{tabular}{|c|c|c|c|c|c|c|c|c|c|} \hline
         \textbf{Inter-} & \multicolumn{3}{c|}{\textbf{i-Vector}} & \multicolumn{3}{c|}{\textbf{X-Vector}} & \multicolumn{3}{c|}{\textbf{DltResNet}} \\ \cline{2-10}
         \textbf{session} & S1 & S2 & S3 & S1 & S2 & S3 & S1 & S2 & S3 \\ \hline
        S1 &  \textbf{5.31} & 11.52 & 10.35 &  \textbf{5.31} & 11.18 & 10.84 &  \textbf{4.85} & 10.69 & 9.56 \\ \hline
        S2 & 11.70 &  \textbf{4.13} & 10.51 & 11.20 &  \textbf{3.51} &  9.96 & 10.63 &  \textbf{4.32} & 9.50 \\ \hline
        S3 & 10.48 & 10.65 &  \textbf{5.16} & 10.70 &  9.96 &  \textbf{5.23} &  9.51 &  9.59 & \textbf{4.53} \\ \hline
    \end{tabular}
\end{table*}

\begin{figure*}[!ht]
    \centering
    \includegraphics[width=0.3\textwidth]{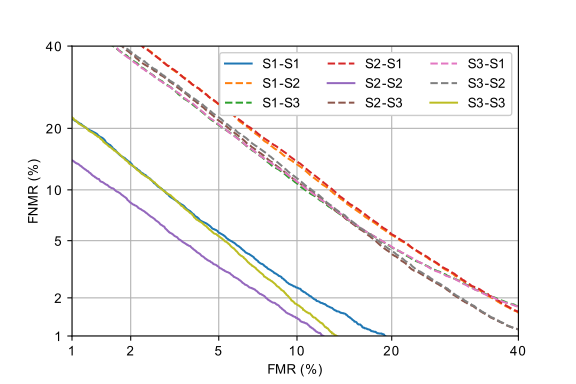}
    \includegraphics[width=0.3\textwidth]{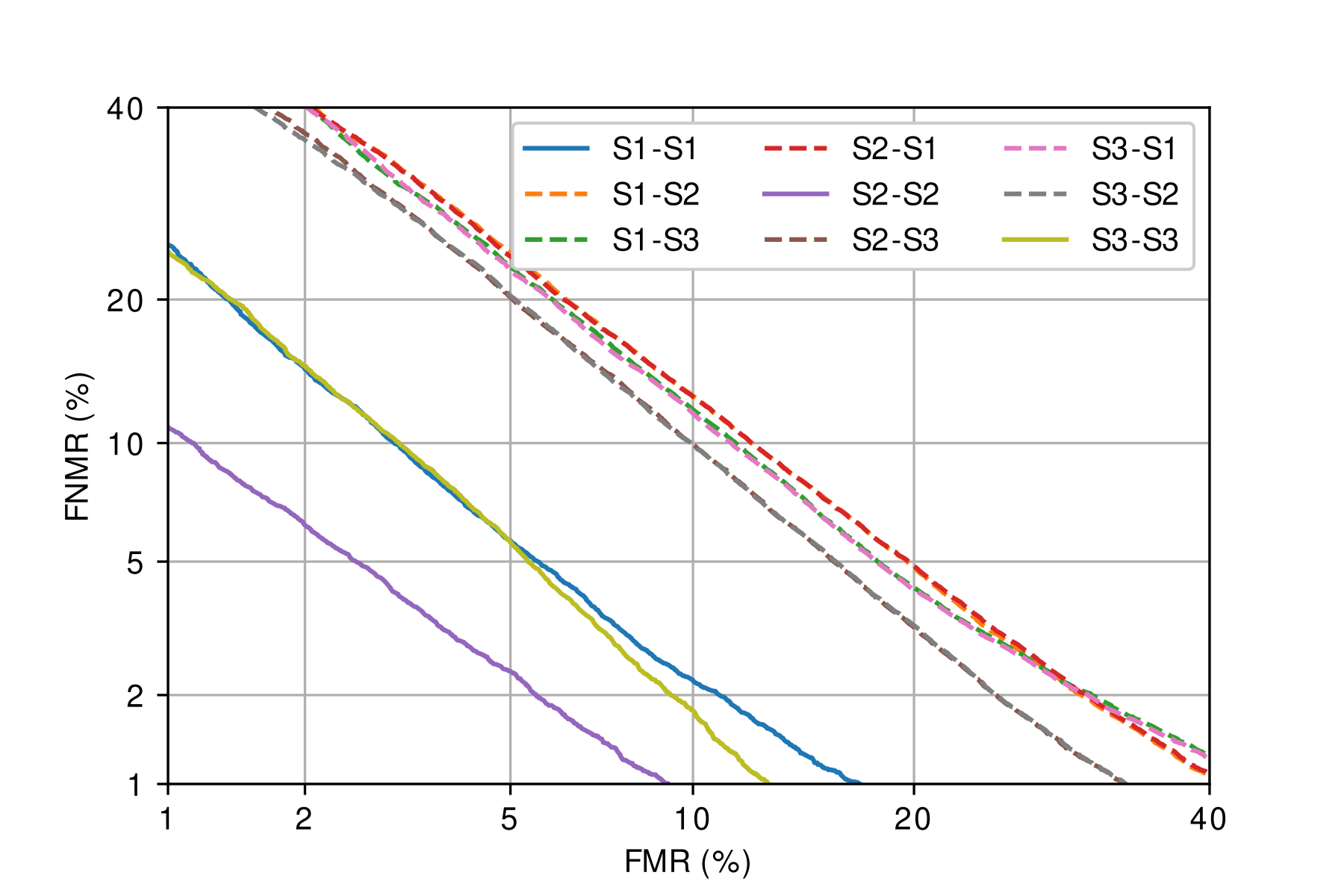}
    \includegraphics[width=0.3\textwidth]{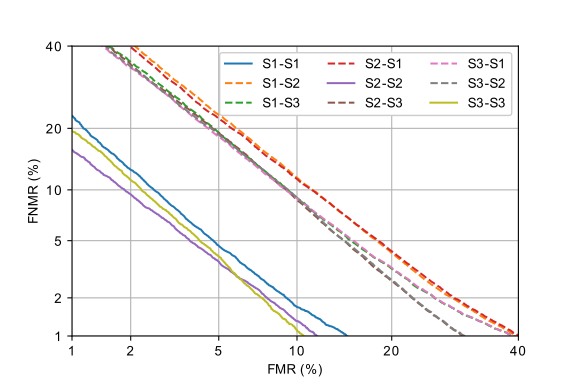}
    \caption{DET curves of inter-session speaker recognition experiments. Left: i-vector, middle: X-vector and right: DltResNet.}
    \label{fig:det_sess_voice}
\end{figure*}

Also, we used Detection Equal Error Rate (D-EER) to present PAD methods' performance, a single value representation of APCER and BPCER. The score distributions of bona fide, zero-effort impostors and attacks are plotted along with the threshold of FMR = 0.1\% to observe the impact of presentation attacks. Detection error trade-off (DET) curves plot the relationship between false match rate (FMR) and false non-match rate (FNMR) for bona fide samples or impostor attack presentation match rate (IAPMR) for attack samples, respectively.

\section{Experimental Results} \label{sec:exp_results}

The main focus of this dataset is to provide scope for developing generalized biometric algorithms in face and speech-based recognition. The generalizability of a biometric algorithm can be achieved by considering multiple dependencies like session variance, device dependency and language. Therefore, in our work, we have performed experiments to demonstrate how these dependencies affect the state-of-the-art face and speaker recognition algorithms mentioned in \ref{sec:perform_eval}. The benchmarking of the dataset is carried out by performing different experiments and presenting the results. 

\subsection{Automatic Speaker Verification}
Automatic Speaker Verification methods display variable performance depending on the channel used to acquire and the noise present in the audio samples. In the following experiments, we have evaluated the performance of the ASV methods in correspondence to the session, device and language. 

\subsubsection{Inter-session speaker recognition}

The MAVS dataset contains data from three different sessions as explained in section \ref{sec:dataset_details}. We have examined the session dependency by performing the inter-session speaker recognition. In this process, we have used the samples from one session to enrol and each of the other sessions to test. Table \ref{tab:audio_session_ee} presents the EER values displaying the comparison of three ASV methods on inter-session experiments.

\begin{itemize}
    \item Session 2 data contains an added noise in all data samples. Therefore, it is seen that higher EER values are observed in all the results where session 2 data is used to enrol.
    \item However, when the same noise is present in test data, the ASV methods tend to perform better than the session with clean data (session 1). This concludes that ASV methods characterize the noise in the data and use it for recognition. 
    \item Similarly, session 3 contains natural noise, which is not consistent in all samples, but it helps recognise the speaker better than the data with no noise. 
    \item Alongside, DltResNet based ASV method displayed better performance compared to other methods.
\end{itemize}

\subsubsection{Inter-device speaker recognition} \label{sec:interdevice_voice}

\begin{table*}[!ht]
    \centering
    \caption{Inter-device speaker recognition evaluation (EER\%) on i-vector method.} \label{tab:ivector_device_eer}
    \begin{tabular}{|c|c|c|c|c|c|} \hline
        \textbf{Inter-device} & iPhone 6s & iPhone 10 & iPhone 11 & Samsung S7 & Samsung S8 \\ \hline
        iPhone 6s & \textbf{1.86} & 5.76 & 6.67 & 15.46 & 14.37 \\ \hline
        iPhone 10 & 5.88 & \textbf{1.62} & 4.74 & 15.02 & 13.97 \\ \hline
        iPhone 11 & 6.73 & 4.67 & \textbf{1.47} & 15.90 & 14.76 \\ \hline
        Samsung S7 & 15.51 & 14.90 & 15.70 & \textbf{10.01} & 13.26 \\ \hline
        Samsung S8 & 14.51 & 13.98 & 14.78 & 13.34 & \textbf{8.77} \\ \hline
    \end{tabular}
\end{table*}

\begin{table*}[!ht]
    \centering
    \caption{Inter-device speaker recognition evaluation (EER\%) on x-vector method.} \label{tab:xvector_device_eer}
    \begin{tabular}{|c|c|c|c|c|c|} \hline
        \textbf{Inter-device} & iPhone 6s & iPhone 10 & iPhone 11 & Samsung S7 & Samsung S8 \\ \hline
        iPhone 6s & \textbf{1.45} & 5.82 & 6.55 & 15.33 & 14.09 \\ \hline
        iPhone 10 & 5.85 & \textbf{1.81} & 4.37 & 13.56 & 12.37 \\ \hline
        iPhone 11 & 6.54 & 4.30 & \textbf{1.81} & 14.27 & 13.10\\ \hline
        Samsung S7 & 15.50 & 13.69 & 14.13 & \textbf{8.55} & 12.97 \\ \hline
        Samsung S8 & 14.04 & 12.25 & 12.93 & 13.30 & \textbf{7.37} \\ \hline
    \end{tabular}
\end{table*}

\begin{table*}[!ht]
    \centering
    \caption{Inter-device speaker recognition evaluation (EER\%) on DltResNet method.} \label{tab:dlrnet_device_eer}
    \begin{tabular}{|c|c|c|c|c|c|} \hline
        \textbf{Inter-device} & iPhone 6s & iPhone 10 & iPhone 11 & Samsung S7 & Samsung S8 \\ \hline
        iPhone 6s & \textbf{2.08} & 6.52 & 7.07 & 16.56 & 16.38\\ \hline
        iPhone 10 & 6.62 & \textbf{2.03} & 4.09 & 15.00 & 15.66 \\ \hline
        iPhone 11 & 7.06 & 4.03 & \textbf{2.02} & 15.92 & 16.14 \\ \hline
        Samsung S7 & 16.68 & 15.07 & 15.83 & \textbf{7.04} & 10.44 \\ \hline
        Samsung S8 & 16.51 & 15.52 & 16.11 & 10.63 & \textbf{7.73} \\ \hline
    \end{tabular}
\end{table*}

The properties of the data capturing device are key attributes for speaker recognition \cite{dehak2011front}. Although state-of-the-art ASV methods accommodate the channel characteristics, the change in devices from enrollment to test can still affect the speaker recognition performance. Our dataset used five different smartphones in data collection to examine the dependency of the device on ASV methods. Tables \ref{tab:ivector_device_eer}, \ref{tab:xvector_device_eer}, \ref{tab:dlrnet_device_eer} show the EERs of all device combinations of enrollment and testing from the three ASV methods.

The results from inter-device experiments output some key points. These observations conclude the impact of channel dependency on state-of-the-art speaker recognition methods. 

\begin{itemize}
    \item The DltResNet method gave out the highest EER in most of the combinations even though it worked better with noisy data as shown in Section \ref{sec:interdevice_voice}. 
    \item The DNN based X-vector methods performed better than other methods. 
    \item It is observed that the combinations of smartphones from the same manufacturer (Apple or Samsung) correlate with speaker recognition. When the enrollment and testing data are from the same manufacturer, the speaker recognition performs better than the cross-manufacturer combination. 
\end{itemize}

\subsubsection{Inter-language speaker recognition} 

\begin{table*}[!ht]
    \centering
        \caption{Inter-language speaker recognition evaluation (EER\%).}  \label{tab:inter_lang_eer}
    \begin{tabular}{|c|c|c|c|c|c|c|c|c|c|} \hline
        \multirow{2}{*}{\textbf{Inter-language}} & \multicolumn{3}{c|}{i-vector} & \multicolumn{3}{c|}{x-vector}& \multicolumn{3}{c|}{DltResNet} \\ \cline{2-10}
         & English & Hindi & Bengali & English & Hindi & Bengali & English & Hindi & Bengali \\ \hline
        English & \textbf{5.47} & 5.50 & 6.72 & \textbf{4.98} & 5.55 & 6.93 & \textbf{4.88} & 5.26 & 6.27 \\ \hline
        Hindi & 5.58 & \textbf{4.16} & 5.33 & 5.45 & \textbf{4.0} & 5.60 & 5.32 & \textbf{3.95} & 5.15 \\ \hline
        Bengali & 6.78 & 5.92 & \textbf{5.08} & 6.93 & 5.67 & \textbf{5.21} & 6.34 & 5.19 & \textbf{4.87} \\ \hline
    \end{tabular}
\end{table*}

\begin{figure*}[!ht]
    \centering
    \includegraphics[width=0.32\textwidth]{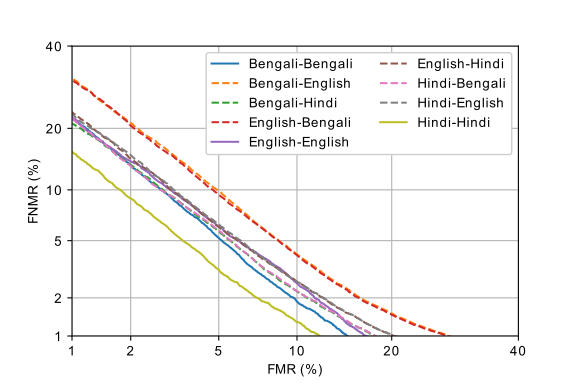}
    \includegraphics[width=0.32\textwidth]{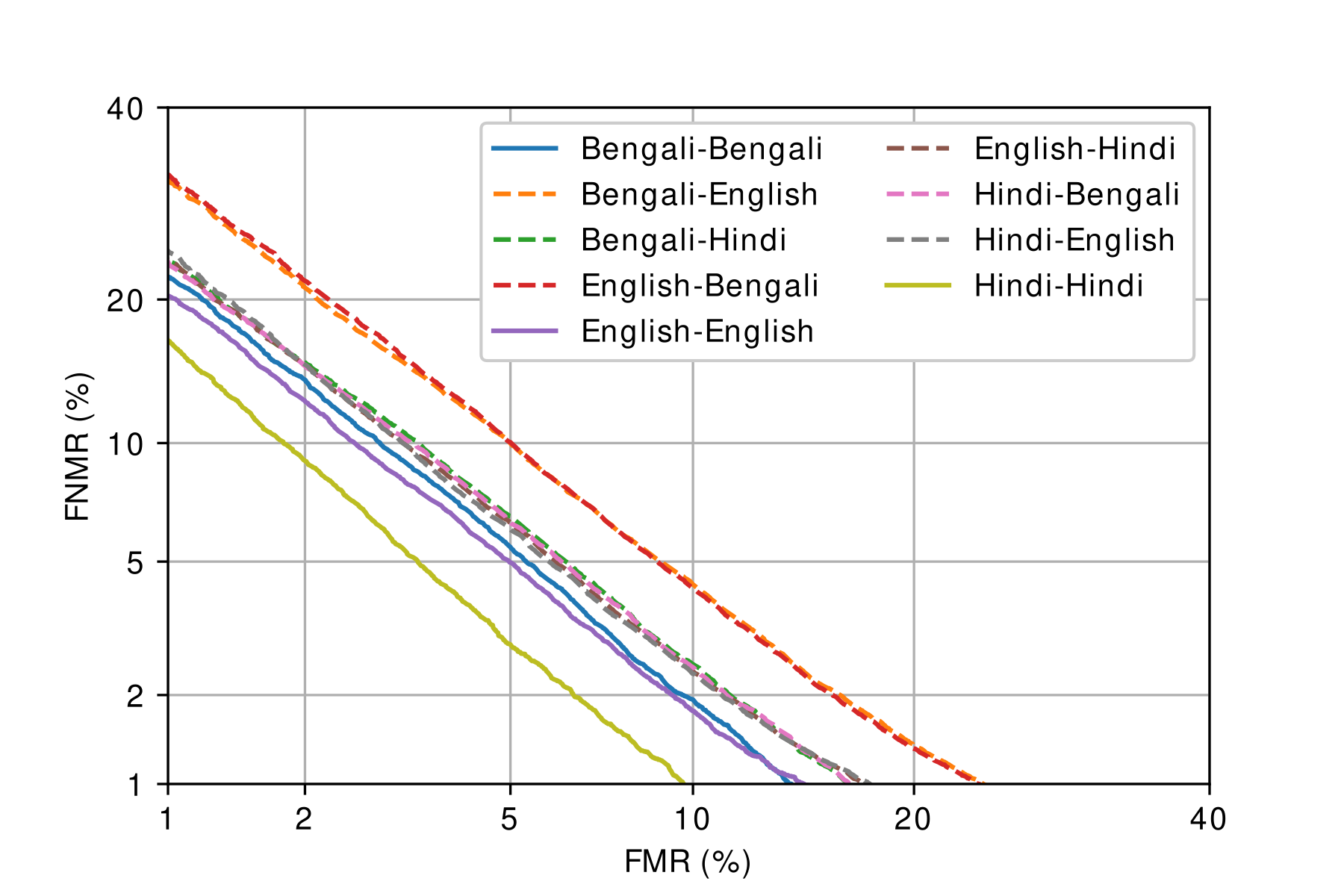}
    \includegraphics[width=0.32\textwidth]{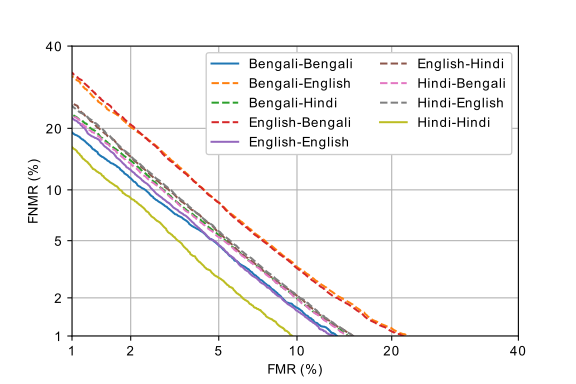}
    \caption{DET curves of inter-language speaker recognition experiments. Left: i-vector, middle: X-vector and right: DltResNet.}
    \label{fig:det_lang_voice}
\end{figure*}

The language difference in the audio sample for ASV has been a hot topic in recent years. Although there are datasets with utterances of the same person in different languages, the problem of language dependency is not benchmarked \cite{ramachandra2019smartphone}. The degradation of biometric recognition due to language mismatch is presented in some previous works \cite{misra2014spoken}, \cite{li2017cross}, \cite{mandalapu2020cross}. Our dataset comprises of the same subjects speaking three different languages, therefore, providing scope for inter-language speaker recognition evaluation. Table \ref{tab:inter_lang_eer} shows the inter-language speaker recognition evaluations. 

\begin{itemize}
    \item The problem of language mismatch from enrollment to testing is observed in all three ASV methods. 
    \item However, the drop in EER is not high, but it is consistent across all the methods. 
    \item It is important to notice that the training dataset contains multiple languages, and we assume that the extracted features contain language factors. 
    \item Therefore, in the scenario of a small subset of languages in training data, the language mismatch problem would be considerable.
\end{itemize}

\subsection{Face Recognition}

\begin{table*}[!ht]
    \centering
    \caption{Inter session face recognition evaluation EER(\%).} \label{tab:face_session_eer}
    \begin{tabular}{|c|c|c|c|c|c|c|c|c|c|} \hline
         \textbf{Inter-} & \multicolumn{3}{c|}{\textbf{LBP}} & \multicolumn{3}{c|}{\textbf{FaceNet}} & \multicolumn{3}{c|}{\textbf{Arcface}} \\ \cline{2-10}
        \textbf{session} & S1 & S2 & S3 & S1 & S2 & S3 & S1 & S2 & S3 \\ \hline
        S1 &  \textbf{5.39} & 24.28 & 44.73 & \textbf{0.26} & 0.89 & 2.22 & \textbf{3.42} & 6.68 & 5.60 \\ \hline
        S2 & 24.28 &  \textbf{6.81} & 41.55 & 0.87 & \textbf{0.24} & 1.65 & 6.42 & \textbf{4.34} & 6.81 \\ \hline
        S3 & 44.67 & 41.43 &  \textbf{4.43} & 2.21 & 1.63 & \textbf{0.21} & 5.59 & 6.81 & \textbf{1.43} \\ \hline
    \end{tabular}
\end{table*}

\begin{figure*}[!ht]
\centering
    \includegraphics[width=0.32\textwidth]{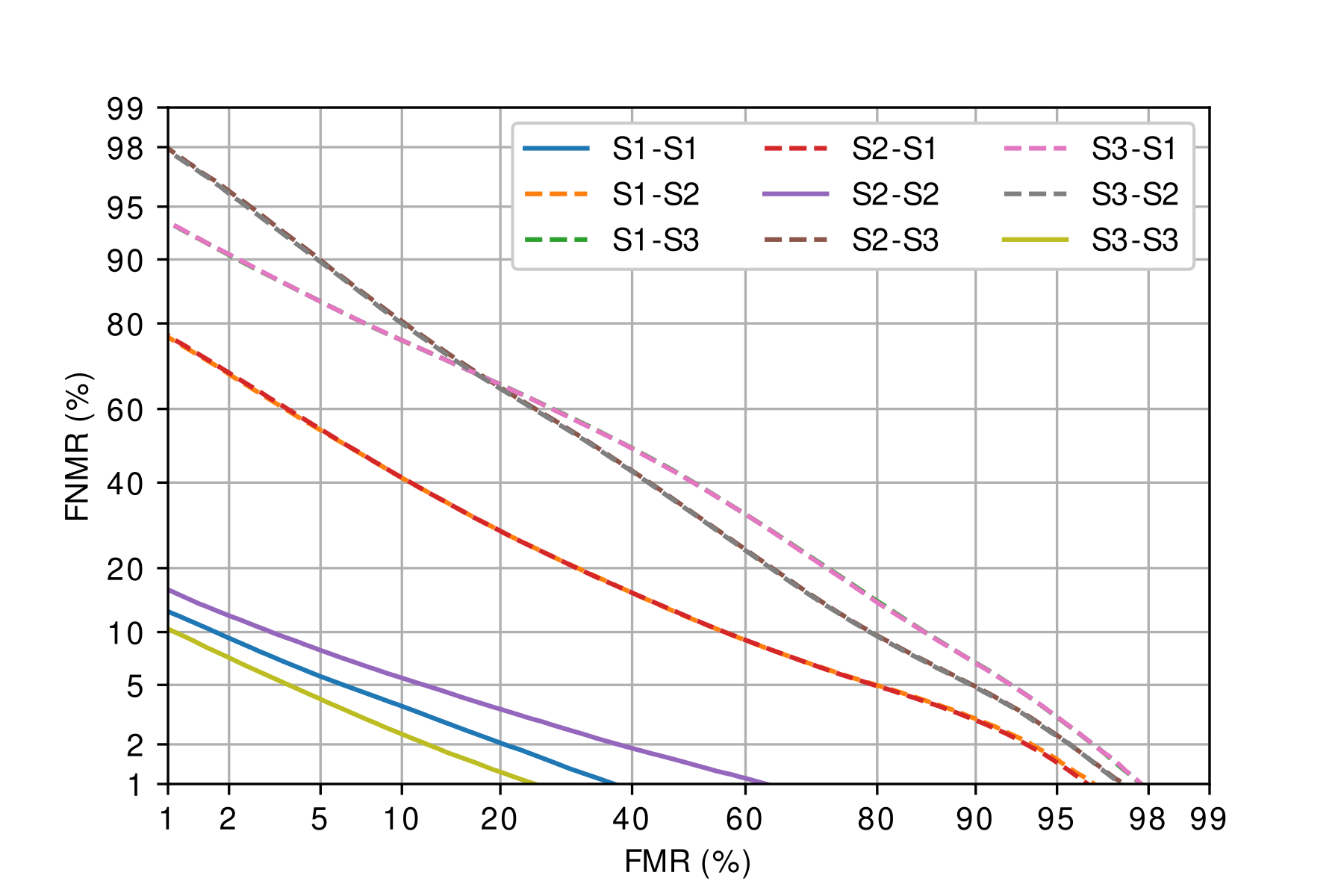}
    \includegraphics[width=0.32\textwidth]{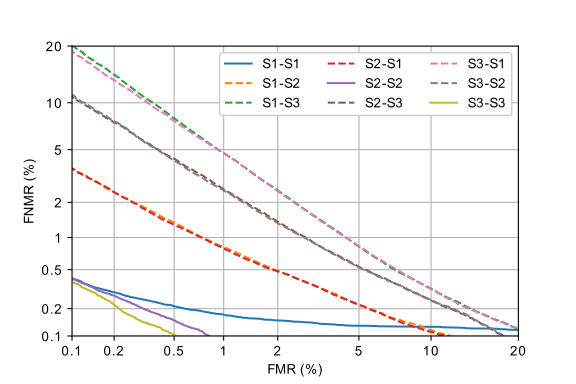}
    \includegraphics[width=0.32\textwidth]{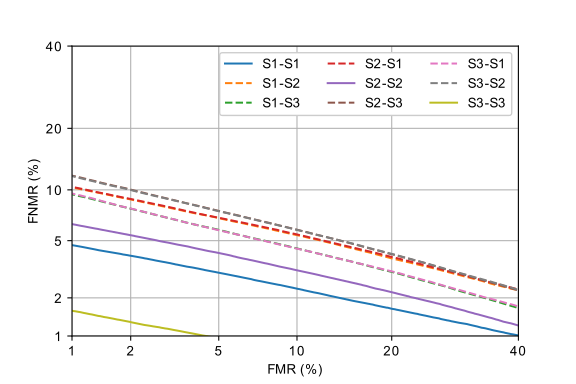}
\caption{DET curves of inter-session face recognition experiments. Left: LBP, middle: FaceNet and right: ArcFace.}
\label{fig:det_sess_face}
\end{figure*}

\begin{table*}[!ht]
    \centering
    \caption{LBP face recognition performance EER(\%) in inter-device scenario.} \label{tab:lbp_device_eer}
    \begin{tabular}{|c|c|c|c|c|c|} \hline
        \textbf{Inter-device} & iPhone 6s & iPhone 10 & iPhone 11 & Samsung S7 & Samsung S8 \\ \hline
        iPhone 6s & \textbf{6.96} & 19.50 & 19.60 & 22.94 & 31.21 \\ \hline
        iPhone 10 & 19.55 & \textbf{5.32} & 18.72 & 31.69 & 37.95 \\ \hline
        iPhone 11 & 19.70 & 18.76 & \textbf{5.09} & 25.67 & 32.60 \\ \hline
        Samsung S7 & 22.96 & 31.69 & 25.70 & \textbf{5.05} & 21.04 \\ \hline
        Samsung S8 & 31.13 & 37.87 & 32.65 & 21.10 & \textbf{5.04} \\ \hline
    \end{tabular}
\end{table*}

\begin{table*}[!ht]
    \centering
    \caption{FaceNet face recognition performance EER(\%) in inter-device scenario.} \label{tab:facenet_device_eer}
    \begin{tabular}{|c|c|c|c|c|c|} \hline
        \textbf{Inter-device} & iPhone 6s & iPhone 10 & iPhone 11 & Samsung S7 & Samsung S8 \\ \hline
        iPhone 6s & \textbf{0.20} & 0.44 & 0.64 & 0.66 & 0.48 \\ \hline
        iPhone 10 & 0.45 & \textbf{0.28} & 0.51 & 0.69 & 0.53 \\ \hline
        iPhone 11 & 0.64 & 0.51 & \textbf{0.3} & 0.92 & 0.71 \\ \hline
        Samsung S7 & 0.67 & 0.68 & 0.90 & \textbf{0.25} & 0.34 \\ \hline
        Samsung S8 & 0.49 & 0.54 & 0.71 & 0.33 & \textbf{0.16} \\ \hline
    \end{tabular}
\end{table*}

\begin{table*}[!ht]
    \centering
    \caption{Arcface face recognition performance EER(\%) in inter-device scenario.} \label{tab:arcface_device_eer}
    \begin{tabular}{|c|c|c|c|c|c|} \hline
        \textbf{Inter-device} & iPhone 6s & iPhone 10 & iPhone 11 & Samsung S7 & Samsung S8 \\ \hline
        iPhone 6s & \textbf{3.30} & 4.14 & 4.03 & 4.79 & 4.36 \\ \hline
        iPhone 10 & 4.10 & \textbf{3.10} & 3.76 & 4.76 & 4.31 \\ \hline
        iPhone 11 & 4.04 & 3.79 & \textbf{3.01} & 4.60 & 4.03 \\ \hline
        Samsung S7 & 4.80 & 4.76 & 4.55 & \textbf{2.98} & 3.78 \\ \hline
        Samsung S8 & 4.39 & 4.30 & 4.03 & 3.78 & \textbf{2.72} \\ \hline
    \end{tabular}
\end{table*}

The robustness of face recognition algorithms in smartphones is evaluated in this section. Similar to speaker recognition, we have performed two dependency experiments, namely inter-session and inter-device. The three face recognition systems are examined in these experiments by taking 20 equally distributed frames in each video. 

\subsubsection{Inter-session}
The session variability in face recognition is observed in this experiment.
\begin{itemize}
    \item  Session 2 and session 3 data has non-uniform lighting on the face region. Therefore, the cross-session face recognition displayed a clear drop in the performance. 
    \item FaceNet performed better in attributing the problem of session variability among the three face recognition methods while displaying near-zero error rates in the same session. 
    \item Table \ref{tab:face_session_eer} present the EER values for inter-session face recognition experiments.
\end{itemize}

\subsubsection{Inter-device}

The results from inter-device experiments on face recognition are shown in Tables \ref{tab:lbp_device_eer}, \ref{tab:facenet_device_eer}, \ref{tab:arcface_device_eer}.

\begin{itemize}
    \item The LBP features based face recognition displayed a high dependency on devices. When the device is the same in enrollment and testing, LBP features performed better face recognition. However, the recognition error has increased by three times when there is a miss-match in devices. 
    \item Another observation is that the change in device manufacturer has also impacted face recognition similar to speaker recognition. 
    \item FaceNet has displayed better face recognition considering the problem of device dependency. The drop in performance is observed, but it is not as consistent as other methods. \item ArcFace performed similarly to FaceNet in an inter-device face recognition scenario. 
    \item Although the EER is higher in ArcFace than FaceNet; the device mismatch has not impacted the performance very much.
\end{itemize} 

\subsection{Audio-Visual Speaker Recognition} \label{sec:av_fusion_method}

The audio-visual speaker recognition is performed by score-level fusion of best-performing face recognition and speaker recognition methods, FaceNet and X-vector methods, respectively. The score fusion approach used in this work is a simple averaging of scores obtained in individual verification methods. 

\subsubsection{Inter-session}

\begin{table}[!ht]
    \centering
    \caption{Inter session Audio-Visual speaker recognition evaluation EER(\%).} \label{tab:score_fusion_session_eer}
    \begin{tabular}{|c|c|c|c|}  \hline
        \textbf{Inter-session} & S1 & S2 & S3  \\ \hline
        S1 &  \textbf{4.99} & 10.73 & 10.46  \\ \hline
        S2 & 10.74 &  \textbf{3.21} &  9.56  \\ \hline
        S3 & 10.34 &  9.55 &  \textbf{4.90} \\ \hline
    \end{tabular}
\end{table}

\begin{itemize}
    \item The combination of audio and visual data displayed similar results as that of individual biometric algorithms. This is because of the simple score-level fusion method employed in our work. 
    \item We assume that an adaptive fusion approach would improve the performance. 
    \item However, it introduces a new dependency on biometric algorithms in the form of a fusion approach. 
    \item Table \ref{tab:score_fusion_session_eer} show the results of inter-session audio-visual fusion experiments. Figure \ref{fig:fusion_sess_det} present the corresponding DET curves.
\end{itemize}

\Figure[tbh]()[width=0.75\columnwidth]{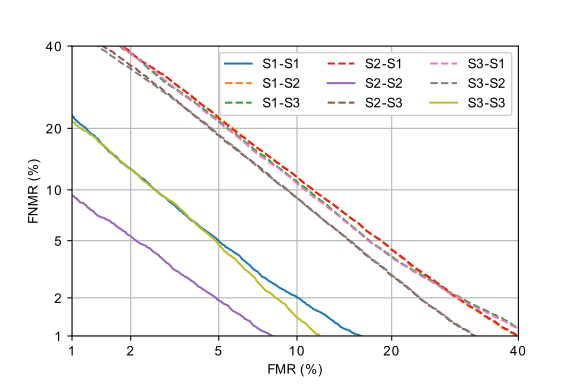}
{DET curves of inter-session experiments on Audio-Visual fusion of FaceNet and X-vector methods. \label{fig:fusion_sess_det}}

\subsubsection{Inter-device}

The inter-device experiments on audio-visual biometric recognition are carried out similar to the inter-session approach. The obtained results display the same observations as that of audio-visual inter-session biometric recognition. It is clear from these experiments that an efficient fusion approach is required to take advantage of bi-modal biometrics. Table \ref{tab:score_fusion_device_eer} display the EER values of inter-device experiments using audio-visual fusion. 

\begin{table*}[htb]
    \centering
    \caption{Inter-device performance (EER\%) of score-level fusion of FaceNet and X-vector methods.} \label{tab:score_fusion_device_eer}
    \begin{tabular}{|c|c|c|c|c|c|} \hline
        \textbf{Inter-device} & iPhone 6s & iPhone 10 & iPhone 11 & Samsung S7 & Samsung S8 \\ \hline
        iPhone 6s & \textbf{1.31} & 5.53 & 6.24 & 14.55 & 13.30 \\ \hline
        iPhone 10 & 5.57 & \textbf{1.65} & 4.18 & 12.82 & 11.74 \\ \hline
        iPhone 11 & 6.25 & 4.15 & \textbf{1.70} & 13.53 & 12.41 \\ \hline
        Samsung S7 & 14.75 & 12.93 & 13.34 & \textbf{7.92} & 12.30 \\ \hline
        Samsung S8 & 13.28 & 11.54 & 12.30 & 12.59 & \textbf{6.81} \\ \hline
    \end{tabular}
\end{table*}

\subsection{Vulnerability from Presentation Attacks} \label{sec:vulnerability}

The vulnerability of biometric recognition towards presentation attacks is examined in this section. The two types of presentation attacks created in this work are explained in Section \ref{sec:pas}. The biometric recognition performance before and after the attacks is compared to check the robustness. When a presentation attack is not carried out, the performance is expressed in false non-match rate (FNMR) caused by zero-effort impostors. In presentation attacks, the vulnerability is presented as impostor attack presentation match rate (IAPMR). 

\subsubsection{Replay Attacks}
The replay attacks are created by replaying an audio-visual biometric sample on a display and loudspeaker combination. The playback sample is recorded on one of the smartphones, namely the Samsung S8. The audio and face channels of replay attacks are examined for vulnerability individually on the two best performed biometric methods from the previous sections. For face recognition, FaceNet features are used, and for speaker recognition, X-vector features are employed.

\begin{itemize}
    \item The impact of replay attack is presented in Table \ref{tab:replay_attack_vuln} in FNMR and IAPMR rates for zero-effort impostors and replay attacks, respectively. 
    \item In face recognition, the vulnerability is observed as 96.87\% IAPMR, representing the number of attacks being matched with bonafide samples. 
    \item The speaker recognition method displayed 25.93\% IAPMR when compared to 6.4\% FNMR. 
    \item The score distributions of bona fide, zero-effort impostors and replay presentation attacks are presented in Figures \ref{fig:voice_replay_attack} and \ref{fig:video_replay_attack}.
\end{itemize}

\begin{table}[tbh]
    \centering
    \caption{Replay attack vulnerability on Face and Voice at FMR = 0.1\%} \label{tab:replay_attack_vuln}
    \begin{tabular}{|c|c|c|} \hline
        \textbf{Biometric} & \textbf{Zero-Effort} & \textbf{Replay} \\ 
        \textbf{Algorithm}& \textbf{impostors} & \textbf{Attacks} \\ \cline{2-3}
         & FNMR & IAPMR \\ \hline
        FaceNet & 0.09\% & 96.87\% \\ \hline
        X-vector & 6.4\% & 25.93\%  \\ \hline
    \end{tabular}
\end{table}

\Figure[htb]()[width=0.96\columnwidth]{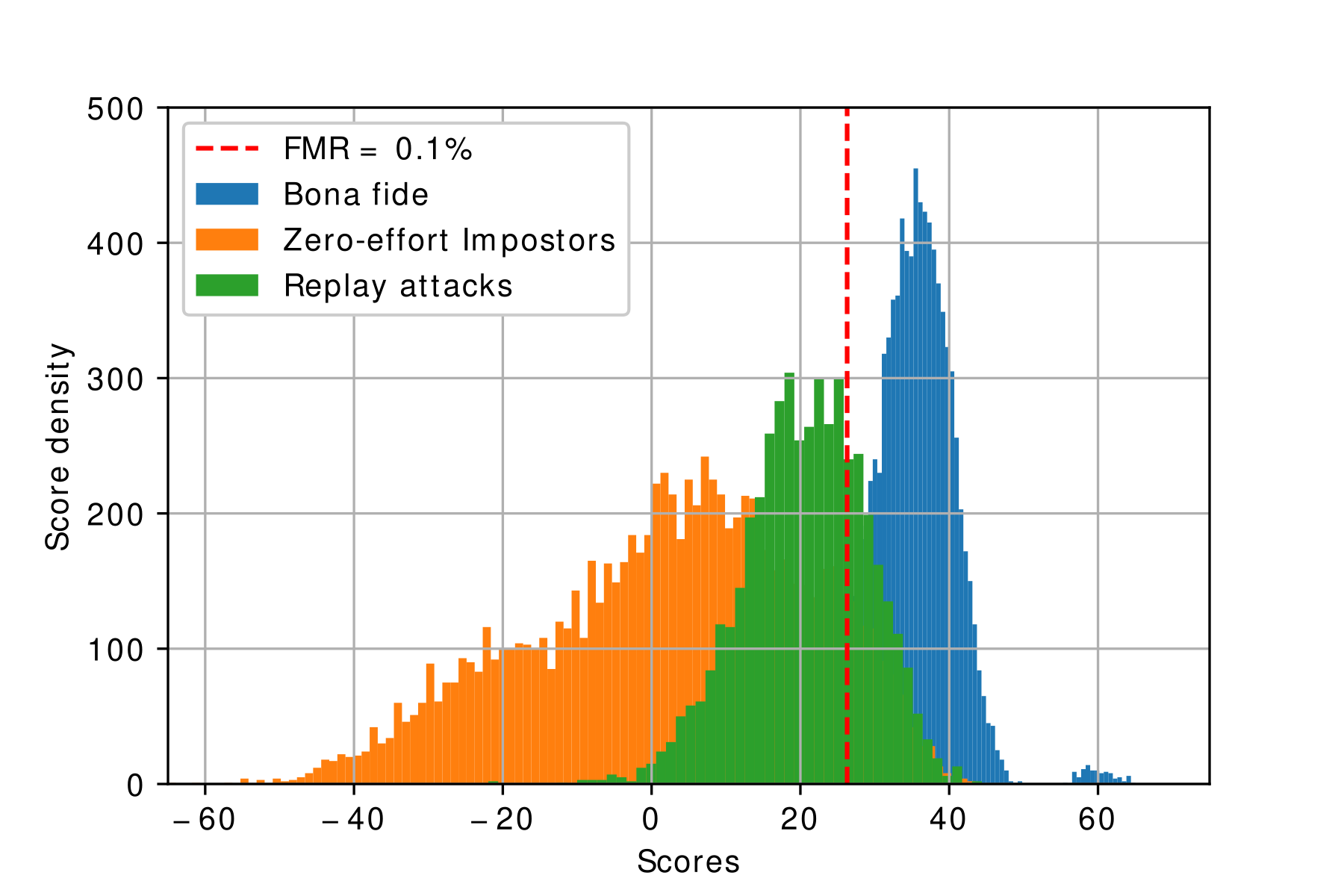}
{Audio Replay attacks score distribution tested on X-vector method.  \label{fig:voice_replay_attack}}

\Figure[htb]()[width=0.96\columnwidth]{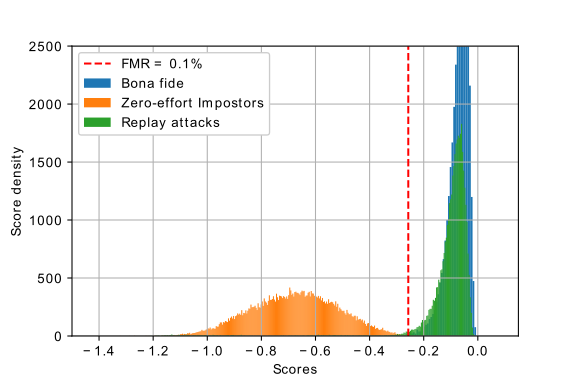}
{Video Replay attacks score distribution tested on FaceNet method.  \label{fig:video_replay_attack}}

\subsubsection{Synthesized Attacks}

Synthesized attacks are logical access attacks where the attack sample is presented digitally to the biometric system. Table \ref{tab:synth_attack_vuln} shows the vulnerability of synthesized attacks on face and voice modalities.

\begin{itemize}
    \item The vulnerability evaluation on FaceNet based face recognition shows a 38.77\% IAPMR, and the score distributions are presented in Figure \ref{fig:video_fsgansource_attack}. 
    \item The speech synthesis is carried out using wavenet-vocoder, and the attacks displayed 99.68\% IAPMR. 
    \item The score distributions are presented in Figure \ref{fig:voice_wavenet_result}.
\end{itemize}


\begin{table}[tbh]
    \centering
    \caption{Synthesized attack vulnerability on Face and Voice at FMR = 0.1\%} \label{tab:synth_attack_vuln}
    \begin{tabular}{|c|c|c|} \hline
        \textbf{Biometric} & \textbf{Zero-Effort} & \textbf{Synthesized} \\ 
        \textbf{Algorithm}& \textbf{impostors} & \textbf{Attacks}\\ \cline{2-3}
         & FNMR & IAPMR \\ \hline 
        FaceNet & 0.21\% & 38.77\% \\ \hline 
        X-vector & 5.59\% & 99.68\% \\ \hline
    \end{tabular}
\end{table}

\Figure[htb]()[width=0.96\columnwidth]{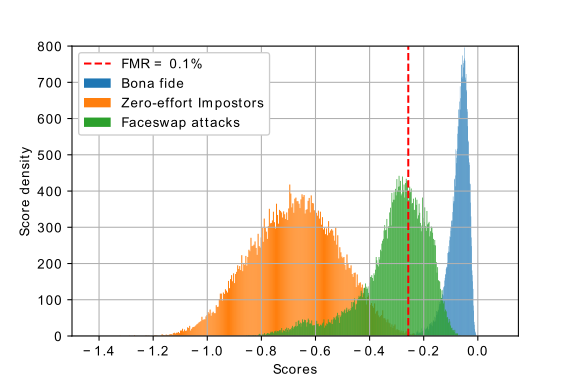}{Score distribution of face swap attacks. \label{fig:video_fsgansource_attack}}


\Figure[htb]()[width=0.96\columnwidth]{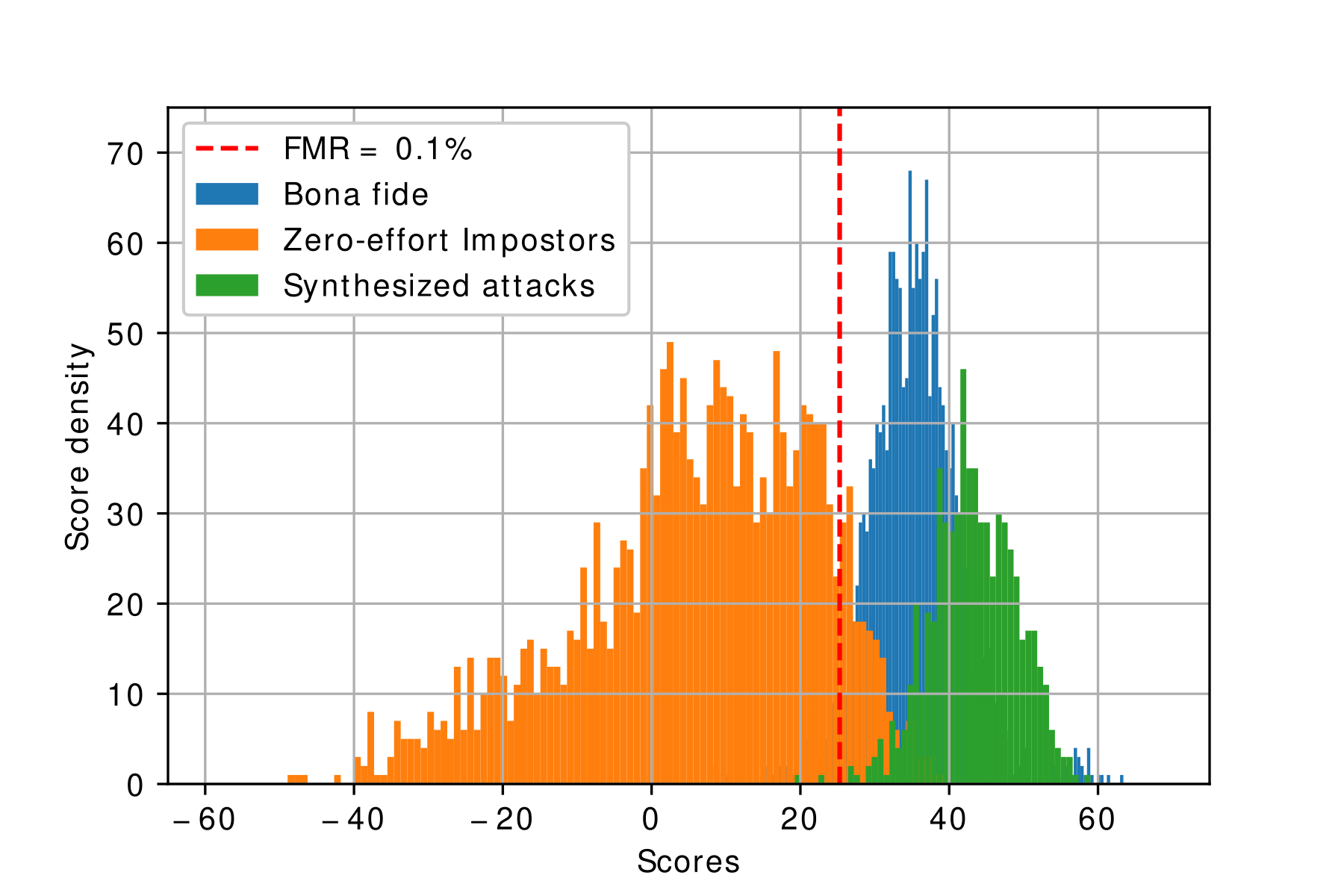}
{Score distributions of wavenet speech synthesized attacks. \label{fig:voice_wavenet_result}}

\subsubsection{Audio-Visual Presentation Attacks}

The vulnerability of audio-visual presentation attacks is examined with the help of fusion of presentation attacks on AV recognition methods explained in Section \ref{sec:av_fusion_method}. The replay attacks and synthesized attacks are performed in individual biometric modalities, and the attack scores are fused to calculate the final scores. The impact of the audio-visual attacks is presented in Table \ref{tab:fusion_attack_score} on two different attacks. Unlike unimodal biometric matching, the results of audio-visual biometrics are presented in False Rejection Rate (FRR) because it represents the system-level performance. Similarly, the score distributions are shown in Figures \ref{fig:fusion_replay_attack_score}, \ref{fig:fusion_synth_attack_scores}.

\begin{table}[tbh]
    \centering
    \caption{Audio-Visual replay attacks vulnerability on AV fusion method at FMR = 0.1\%} \label{tab:fusion_attack_score}
    \begin{tabular}{|c|c|c|} \hline
        \textbf{Attack} & \textbf{Zero-Effort} & \textbf{Presentation} \\ 
        \textbf{Type}& \textbf{impostors} & \textbf{Attacks}\\ \cline{2-3}
         & FNMR & IAPMR \\ \hline 
        Replay Attacks & 5.29\% & 28.46\% \\ \hline 
        Synthesized Attacks & 4.64\% & 99.83\% \\ \hline
    \end{tabular}
\end{table}

\Figure[htb]()[width=0.96\columnwidth]{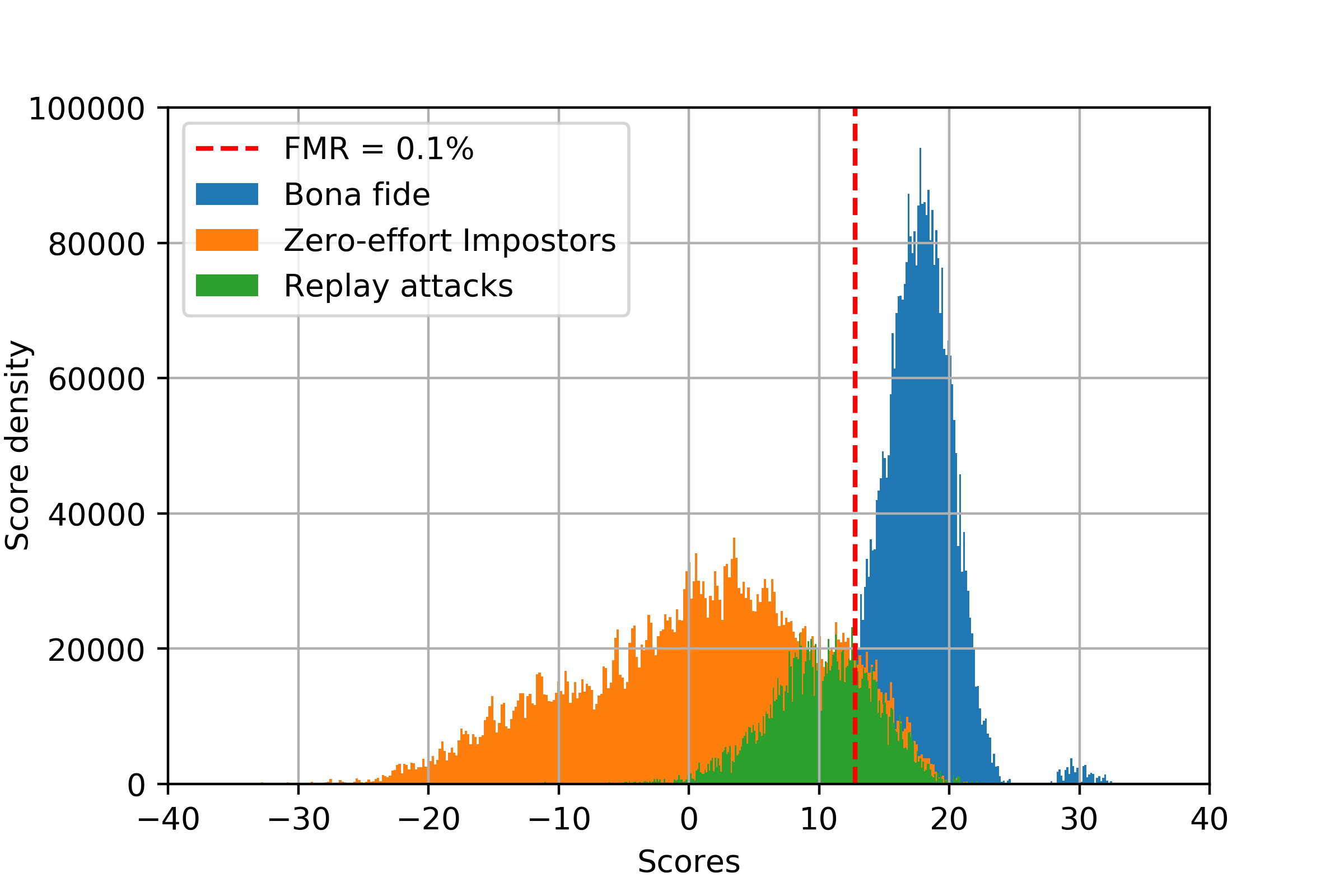}{Audio-Visual replay attacks score distribution. \label{fig:fusion_replay_attack_score}}

\Figure[htb]()[width=0.96\columnwidth]{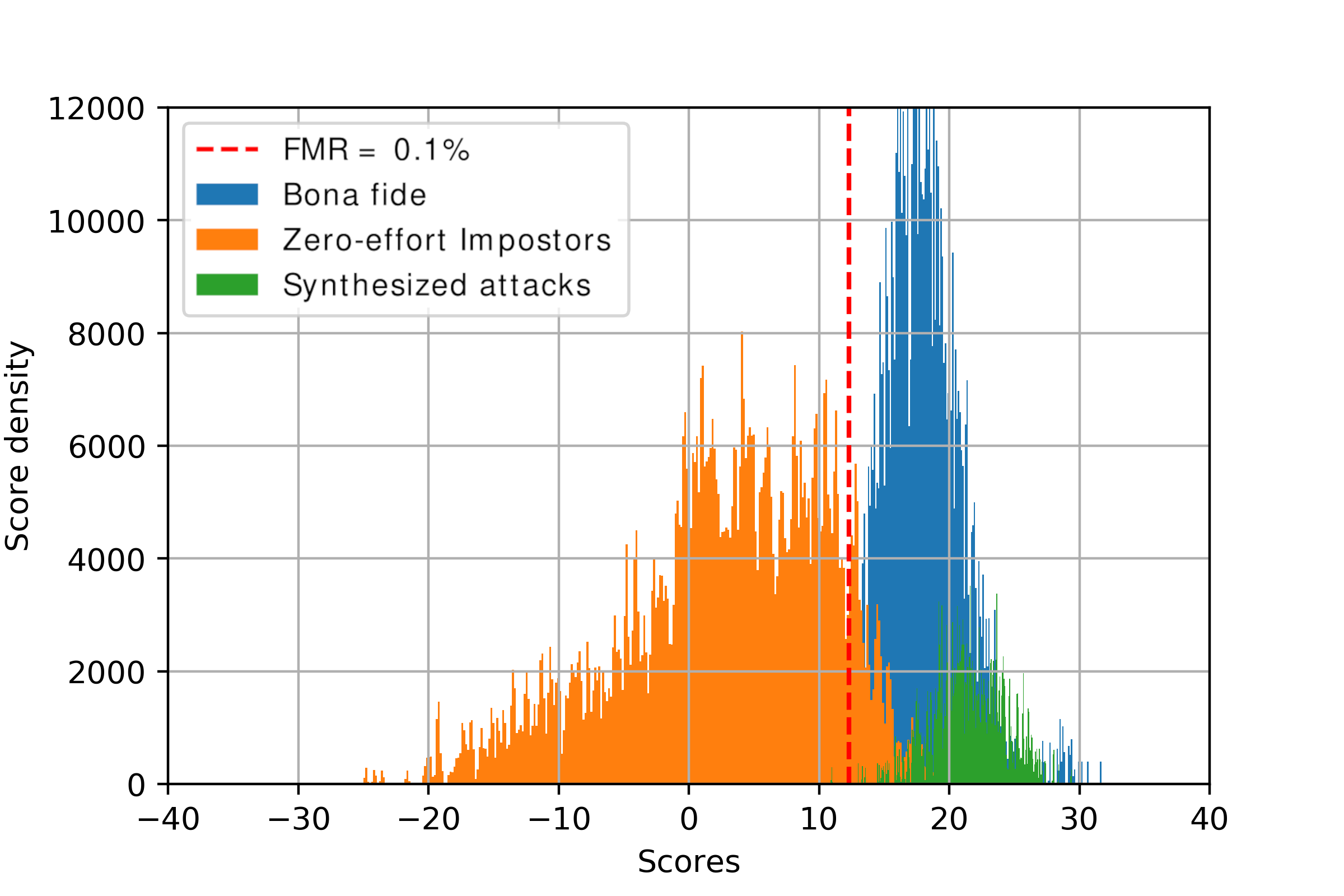}{Audio-Visual synthesized attacks score distribution. \label{fig:fusion_synth_attack_scores}}

\begin{itemize}
    \item The results indicate that audio-visual fusion is vulnerable to presentation attacks. 
    \item The problem of replay attacks is less compared to the synthesized attacks. 
    \item Although the replay attacks on face recognition displayed the highest vulnerability; the AV fusion approach appears to have the ability to overcome this problem. However, a similar observation is not seen in synthesized attacks. 
    \item Thus, the AV fusion recognition approach has the vulnerability due to combined AV presentation attacks. 
\end{itemize}

\subsection{Presentation Attack Detection} \label{sec:pad_results}

\begin{table*}[tbh]
    \centering
    \caption{Results of speaker recognition presentation attack detection.} \label{tab:voice_pad}
    \begin{tabular}{|c|c|c|c|c|c|c|} \hline
        \textbf{Attack} & \multicolumn{3}{c|}{LFCC-GMM} & \multicolumn{3}{c|}{CQCC-GMM} \\ \cline{2-7}
        \textbf{type} & D-EER & BPCER\_5 & BPCER\_10 & D-EER & BPCER\_5 & BPCER\_10 \\ \hline
        \textbf{Replay Attacks} & 44.14\% & 100\% & 93.15\% & 20.49\% & 45.63\% & 36.89\% \\ \hline
        \textbf{Speech Synthesis} & 14.00\% & 39.82\% & 20.38\% & 14.08\% & 40.77\% & 22.33\% \\ \hline
    \end{tabular}
\end{table*}

\begin{table*}[tbh]
    \centering
    \caption{Results of face recognition presentation attack detection.} \label{tab:face_pad}
    \begin{tabular}{|c|c|c|c|c|c|c|} \hline
        \textbf{Attack} & \multicolumn{3}{c|}{LBP-SVM} & \multicolumn{3}{c|}{Color texture-SVM} \\ \cline{2-7}
        \textbf{type} & D-EER & BPCER\_5 & BPCER\_10 & D-EER & BPCER\_5 & BPCER\_10 \\ \hline
        \textbf{Replay Attacks} & 4.96\% & 5.07\% & 1.28\% & 2.15\% & 1.35\% & 0.32\% \\ \hline
        \textbf{FaceSwap} & 2.99\% & 1.74\% & 1.15\% & 2.54\% & 0.83\% & 0.26\% \\ \hline
    \end{tabular}
\end{table*}

The presentation attack detection experiments are performed using baseline PAD methods. The attack data is partitioned into three sets: training, developing and testing, with 35\%, 35\% and 30\% of bona fide and attack samples, respectively. Each partition includes data from a unique set of subjects. We have chosen the baseline approaches used in Automatic Speaker Verification Spoofing and Countermeasures Challenge (ASVSpoof) for speaker recognition PAD in 2019. See Section \ref{sec:pad_methods}. For face recognition, we opted the two best-performing methods from the face PAD methods used in \cite{ramachandra2019smartphone}. Tables \ref{tab:voice_pad} and \ref{tab:face_pad} show the results of the PAD methods in terms of D-EER, BPCER at APCER = 5\% and BPCER at APCER = 10\%. The DET curves in figures \ref{fig:voice_pad} and \ref{fig:face_pad} present the performance of PAD methods.

\Figure[tbh]()[width=0.96\columnwidth]{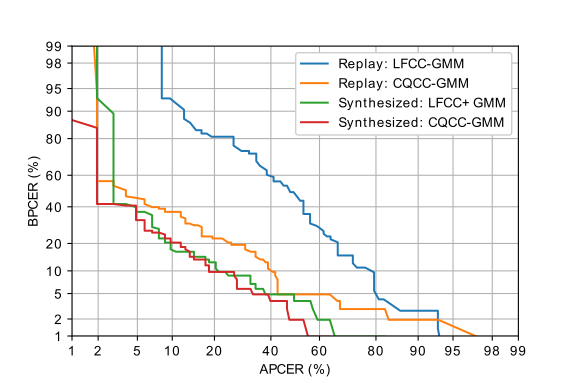}
{DET curves of voice PAD evaluation using baseline methods. \label{fig:voice_pad}}

\Figure[tbh]()[width=0.96\columnwidth]{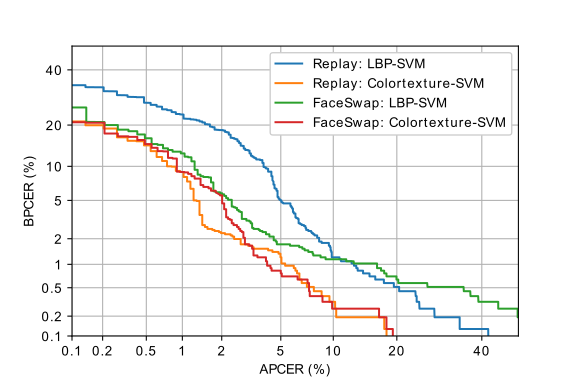}
{DET curves of face PAD evaluation using baseline methods. \label{fig:face_pad}}

\begin{itemize}
    \item The voice PAD results indicate that the baseline methods are not able to detect the attacks.
    \item Alongside, replay attacks are difficult to detect when compared to synthesized attacks. In contrast, both face PAD methods performed well in detecting the attacks. 
    \item The voice PAD methods are tested on the whole speech sample, where the face PAD methods are performed on detected face images in individual frames. 
    \item Therefore, it is reasonable to assume that this could be the reason for the difference in performance. 
\end{itemize}

\subsubsection{Multimodal PAD}

The presentation attacks on both modalities are possible with sophisticated equipment. The PAD methods should be able to detect the attacks before the verification process. In this experiment, we have fused the PAD scores from the CQCC-GMM method and the Color texture-SVM method to compute multimodal PAD scores. We have used a sum rule based fusion to combine two PAD methods. The table \ref{tab:fusion_pad} shows the results of multimodal PAD approach and Figure \ref{fig:fusion_pad} shows the PAD performance on two different types of attacks.

\begin{table}[tbh]
    \centering
    \caption{Results of audio-visual PAD methods.} \label{tab:fusion_pad}
    \begin{tabular}{|c|c|c|c|} \hline
        \textbf{Attack} & \multicolumn{3}{c|}{Fusion PAD} \\ \cline{2-4}
        \textbf{type} & D-EER & BPCER\_5 & BPCER\_10  \\ \hline
        \textbf{Replay Attacks} & 16.99\% & 38.83\% & 30.10\% \\ \hline
        \textbf{Synthesized} & 11.87\% & 32.04\% & 15.54\%  \\ \hline
    \end{tabular}
\end{table}

\begin{itemize}
    \item The replay attacks are observed to be difficult to detect compared to synthesized attacks. The performance of multimodal PAD is similar to individual PAD in regards to the types of attacks.
    \item The multimodal PAD does not improve the attack detection performance. The reason for this could be the usage of simple sum rule based fusion.
    \item The co-related and complementary information between audio and visual domains is not taken into account in this fusion approach. Therefore, multimodal PAD does not show any promising improvement over individual PAD approaches.
\end{itemize}

\Figure[tbh]()[width=0.96\columnwidth]{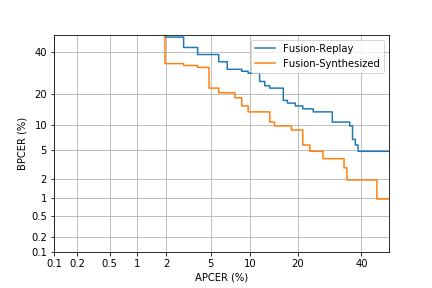}
{DET curves of audio-visual PAD of CQCC and Color texture methods. \label{fig:fusion_pad}.}

\section{Conclusion} \label{sec:conclusion}

Smartphone biometrics have emerged into advanced security applications like banking transactions and identity verification. The built-in biometric systems by smartphone manufacturers can be utilized for this purpose. However, it is difficult to entirely rely on the built-in systems due to the variance in sensors and unknown algorithms embedded into smartphones. In this direction, it is possible to use the default sensors in smartphones like cameras and microphones. Therefore, we have developed a multidimensional smartphone audio-visual dataset that includes different languages, devices, sessions, and texts in this work. We have presented in this paper some of the previous works on building an audio-visual dataset and discussed our multi-lingual smartphone audio-visual (MAVS) dataset. 

Further, we have performed experiments on examining the robustness of state-of-the-art biometric algorithms in two directions. The first direction concerns the problem of algorithm dependencies that include signal noise, capturing device and speech language. We have prepared inter-session, inter-device and inter-language experiments and presented the results. In the second direction, presentation attacks are evaluated for the vulnerability of biometric algorithms and the performance of baseline PAD algorithms. The results show the requirement of robust audio-visual biometrics algorithms to deal with the problems of multiple dependencies and presentation attacks. The proposed dataset would help the research community in developing advanced biometric algorithms and presentation attack detection approaches.

\subsection{Future work}

The MAVS dataset is made publicly available for research purposes \footnote{MAVS dataset request form: \url{https://docs.google.com/forms/d/e/1FAIpQLSfTMqnQj8KNoUi1Ms1tx8Ewgil2l4wAAJVaKUJs6VkWfjAo4w/viewform?usp=sf_link}}. The proposed dataset can be used in multiple directions in smartphone audio-visual research. The future work in this research direction using the dataset is as follows.

\begin{enumerate}
    \item Novel biometric algorithms are modelled by identifying various problems that question the robustness of smartphone authentication.
    \item The authentication technology through biometrics can be improved via Audio-visual person recognition through the efficient usage of complementary information between audio and visual modalities.
    \item The dataset contains subjects of different ages ranging from 18 to 48 years and gender labels (70 male and 33 female). Therefore, the dataset can be used for studying gender classification and fairness. Further, the audio data from three different languages can be used for language detection.
    \item The correlated information between biometric cues are used to propose advanced presentation attack detection algorithms towards unknown and unseen attacks. E.g. lip-sync, correlated biometric data.
    \item Generalizable biometric algorithms are developed in smartphone environments for real-world applications across different devices and capturing conditions.
\end{enumerate}

\section*{Acknowledgment}

We acknowledge the Idiap Research Institute and Prof. Sébastien Marcel for the data capture mobile application developed as a part of the SWAN (Secured access over Wide Area Network) project funded by the Research Council of Norway (Grant No. IKTPLUSS 248030/O70).

\bibliography{references.bib}
\bibliographystyle{plain}
    
\begin{IEEEbiography}[{\includegraphics[width=1in,height=1.25in,clip,keepaspectratio]{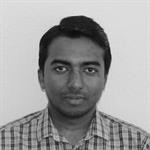}}]{Hareesh Mandalapu} received M.Tech. degree in Computer Science from University of Hyderabad in 2015 and M.S. in Erasmus Masters CIMET from Université Jean Monnet, France in 2017. He is currently pursuing the Ph.D. degree in Information Security and Communication Technology from Norwegian University of Science and Technology, Gjøvik, Norway. His research interests include audio-visual biometrics, presentation attack detection and multilingual speaker recognition. \end{IEEEbiography}

\begin{IEEEbiography}[{\includegraphics[width=1in,height=1.25in,clip,keepaspectratio]{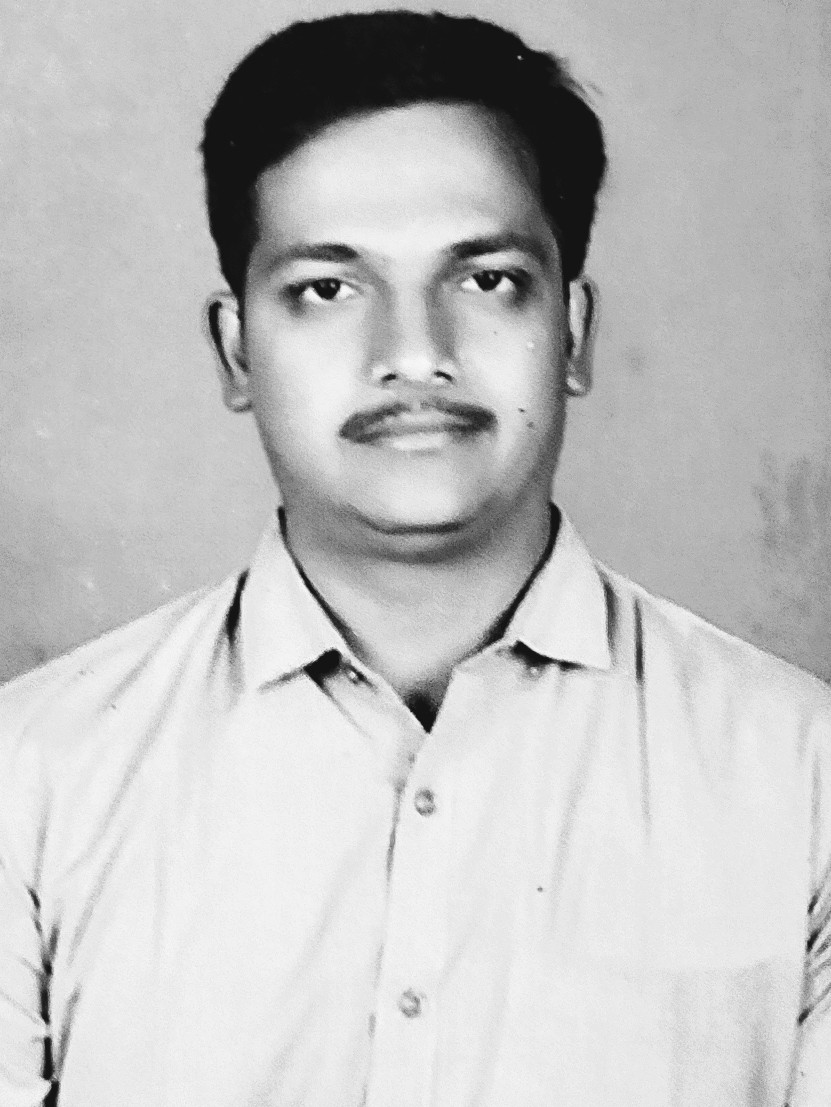}}]{Aravinda Reddy P N} received M.Tech. degree in Signal Processing from Visvesvaraya Technological University Belgaum in 2014. He is currently pursuing Ph.D. degree in Advanced Technology Development Centre, Indian Institute of Technology Kharagpur, Kharagpur, West Bengal, India. His research interests include automatic speech recognition, audio-visual biometrics, presentation attack detection. \end{IEEEbiography}

\begin{IEEEbiography}[{\includegraphics[width=1in,height=1.25in,clip,keepaspectratio]{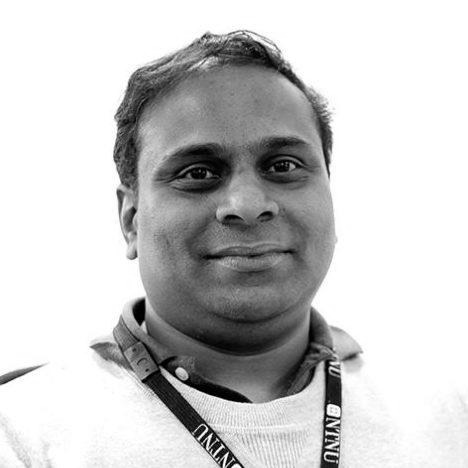}}]{Raghavendra Ramachandra} obtained his Ph.D. in Computer Science and Technology from the University of Mysore, Mysore India and Institute Telecom, and Telecom Sudparis, Evry, France (carried out as a collaborative work) in 2010. He is currently appointed as a Full Professor with the Institute of Information Security and communication technology (IIK), Norwegian University of Science and Technology (NTNU), Gjøvik, Norway. He was a Researcher with the Istituto Italiano di Tecnologia, Genoa, Italy, where he worked with video surveillance and social signal processing. His main research interests include Deep Learning, statistical pattern recognition, data fusion schemes, and random optimization, with applications to biometrics, multimodal biometric fusion, human behaviour analysis, and crowd behaviour analysis. He has authored several papers and is a reviewer for several international conferences and journals. He also holds several patents in biometric presentation attack detection.  He was/is also involved in various conference organizing and program committees and serving as an associate editor for various journals. He was/is participating (as PI/Co-PI/contributor) in several EU projects, IARPA USA and other national projects.  He has served as an editor for ISO/IEC 24722 standards on multimodal biometrics and an active contributor for ISO/IEC SC 37 standards on biometrics.  He has received several best paper awards, and he is also a senior member of IEEE. \end{IEEEbiography}

\begin{IEEEbiography}[{\includegraphics[width=1in,height=1.25in,clip,keepaspectratio]{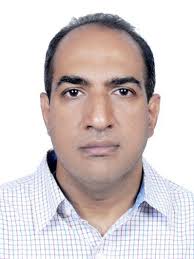}}]{Krothapalli Sreenivasa Rao} received the B.Tech degree in Electronics and communication from RVR college of engineering in 1990 and received M.E degree in Communication Systems from PSG Tech, Coimbatore, India, in 2006. and received PhD from Dept of Computer Science and Engineering, IIT Madras, Chennai, India in 2004. Currently, he is working as Professor in Department of Computer Science and Engineering, IIT Kharagpur, Kharagpur, West Bengal, India. He has supervised 7 PhDs and 14 MS (by research) in different issues related to speech processing. \end{IEEEbiography}

\begin{IEEEbiography}[{\includegraphics[width=1in,height=1.25in,clip,keepaspectratio]{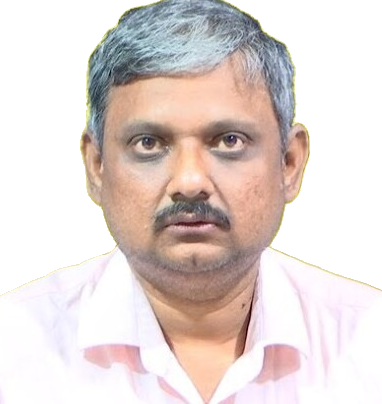}}]{Pabitra Mitra} received the B.Tech degree Electrical Engineering, IIT Kharagpur, Kharagpur, India in 1996 and received PhD from Dept Computer Science and Engineering, Indian Statistical Institute, Kolkata, India in 2005. Currently, he is working as Professor in Department of Computer Science and Engineering, IIT Kharagpur, Kharagpur, West Bengal, India. He has  supervised 8 PhDs and 12 MS (by research) in different issues related to AI and Machine learning. \end{IEEEbiography}

\begin{IEEEbiography}[{\includegraphics[width=1in,height=1.25in,clip,keepaspectratio]{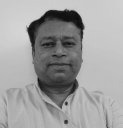}}]{S.R. Mahadeva Prasanna} received the B.Tech degree in Electronics and communication from SSIT Tumakuru, Tumakuru, Karnataka, India in 1994 and received M.Tech degree in Industrial Electronics from NIT Surathkal, Surathkal,Karnataka, India in 1997 and received PhD from Dept of Computer Science and Engineering, IIT Madras, Chennai, India in 2004. Currently, he is working as Professor in Department of Electrical Engineering, IIT Dharwad, Dharwad, Karnataka, India. He has supervised 13 PhDs in different issues related to speech processing. \end{IEEEbiography}


\begin{IEEEbiography}[{\includegraphics[width=1in,height=1.25in,clip,keepaspectratio]{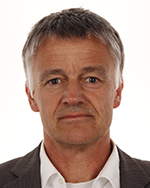}}]{Christoph Busch} (Senior Member, IEEE) is a member of the Department of Information Security and Communication Technology (IIK), Norwegian University of Science and Technology (NTNU), Norway. He holds a joint appointment with the Faculty of Computer Science, Hochschule Darmstadt (HDA), Germany. Furthermore, he has been a Lecturer of biometric systems with the Technical University of Denmark (DTU), since 2007. He coauthored more than 400 technical papers and has been a speaker at international conferences. He is a convenor of WG3 in ISO/IEC JTC1 SC37 on biometrics and an active member of CEN TC 224 WG18. He served for various program committees, such as NIST IBPC, ICB, ICHB, BSI-Congress, GI-Congress, DACH, WEDELMUSIC, and EUROGRAPHICS, and served for several conferences, journals, and magazines as a Reviewer such as ACM-SIGGRAPH, ACM-TISSEC, the IEEE Computer Graphics and Applications, the IEEE Transactions on Signal Processing, the IEEE Transactions on Information Forensics and Security, the IEEE Transactions on Pattern Analysis and Machine Intelligence, and the Computers and Security journal (Elsevier). Furthermore, on behalf of Fraunhofer, he chairs the biometrics working group of the TeleTrusT association as well as the German standardization body on biometrics (DIN-NIA37). He is also an Appointed Member of the Editorial Board of the IET Biometrics journal and the IEEE Transactions on Information Forensics and Security journal. \end{IEEEbiography}

\EOD

\end{document}